\documentclass[acmsmall,screen]{acmart}

\AtBeginDocument{
  }

\usepackage[latin1]{inputenc}
\usepackage{graphicx}
\usepackage{amsmath}
\usepackage{amsthm}
\usepackage{url}
\usepackage{epic,eepic,latexsym,color}
\usepackage{ifthen,graphics,epsfig}
\usepackage{wrapfig}
\usepackage{xspace}
\usepackage{color}

\newtheorem{theorem}{Theorem}[section]
\newtheorem{lemma}{Lemma}[section]
\newtheorem{corollary}{Corollary}[section]
\newtheorem{definition}{Definition}[section]
\newtheorem{remark}{Remark}[section]
\newtheorem{claim}{Claim}[section]

\newcounter{linecounter}
\newcommand{\linenumbering}{\ifthenelse{\value{linecounter}<10}{(0\arabic{linecounter})}{(\arabic{linecounter})}}
\renewcommand{\line}[1]{\refstepcounter{linecounter}\label{#1}\linenumbering}
\newcommand{\resetline}[1]{\setcounter{linecounter}{0}#1}
\renewcommand{\thelinecounter}{\ifnum \value{linecounter} >
9\else 0\fi \arabic{linecounter}}

\newcommand{\remove}[1]{}

\newcommand {\CAS} {{\sf Compare\&Swap}\xspace}
\newcommand {\RMW} {{\sf Read-Modify-Write}\xspace}

\newcommand {\FAI} {{\sf Fetch\&Inc}\xspace}

\newcommand {\R} {{\sf Read}\xspace}
\newcommand {\W} {{\sf Write}\xspace}
\newcommand {\Snap} {{\sf Snapshot}\xspace}

\newcommand {\Pop} {{\sf Pop}\xspace}
\newcommand {\Push} {{\sf Push}\xspace}

\newcommand {\Enq} {{\sf Enqueue}\xspace}
\newcommand {\Deq} {{\sf Dequeue}\xspace}

\newcommand {\error} {{\sf ERROR}\xspace}

\newcommand {\Verify} {{\sf Verify}\xspace}
\newcommand {\Apply} {{\sf Apply}\xspace}
\newcommand {\RV} {{\sf DRV}\xspace}
\newcommand {\GenLin} {{\sf GenLin}\xspace}

\newcommand{\m}[1]{\ensuremath{\mathcal{#1}}\xspace}

\newcommand{\op}{\mbox{\sl op}}

\setcopyright{acmcopyright}
\copyrightyear{2025}
\acmYear{2025}
\acmDOI{XXXXXXX.XXXXXXX}


\begin{document}

\title{Asynchronous Wait-Free Runtime Verification and Enforcement of Linearizability}

\thanks{$\star$ The conference version of this paper can be found in~\cite{CR23}.}

\author{Armando Casta\~neda}
\affiliation{
  \institution{Instituto de Matem\'aticas, Universdad Nacional Aut\'onoma de M\'exico}
  \city{Mexico City}
  \country{Mexico}
}
\orcid{https://orcid.org/0000-0002-8017-8639}
\email{armando.castaneda@im.unam.mx}

\author{Gilde Valeria Rodr\'iguez}
\affiliation{
  \institution{Posgrado en Ciencia e Ingenier\'ia de la Computaci\'on,
      Universidad Nacional Aut\'onoma de M\'exico}
  \city{Mexico City}
  \country{Mexico}
  }
\orcid{https://orcid.org/0009-0009-1463-7786}
\email{gilde@ciencias.unam.mx}

\renewcommand{\shortauthors}{Casta\~neda and Rodr\'iguez}

\begin{abstract}
This paper presents a {theoretical study} of the problem of verifying
linearizability at runtime, where 
one seeks for a concurrent algorithm for verifying that the current execution
of a given concurrent shared object implementation is linearizable.
It shows that it is impossible to runtime verify linearizability for some common 
sequential objects, regardless of the consensus power of base objects.
Then, it argues that a variant of the problem, which we call predictive verification, can be solved,
if linearizability is verified indirectly. Namely, it shows that
(1)~linearizability of a class of concurrent implementations can be predictively verified
using only read/write base objects (i.e. without the need of consensus), and 
(2)~any implementation can be transformed to its counterpart in the class 
using only read/write objects.
As far as we know, this is the first runtime verification algorithm for any correctness condition
that is fully asynchronous and fault-tolerant. 
{As a by-product, it is obtained a simple and generic methodology for deriving  
linearizable implementations that runtime verify their responses, and 
are able to produce a history certifying this, properties that allows the design of 
concurrent systems in a modular manner
with accountable and forensic guarantees.
We call such implementations self-enforced linearizable.}
The results hold not only for linearizability but
for a correctness condition that includes generalizations
of it such as set-linearizability and interval-linearizability.
\end{abstract}


\maketitle

\section{Introduction}
\label{sec-intro}

\subsection{Linearizability and its challenges}

\emph{Linearizability}~\cite{HW90} is the de facto
correctness condition for asynchronous shared memory 
concurrent implementations of objects defined through
sequential specifications.
Intuitively, an implementation is linearizable if each operation
happens atomically at a single moment of time between its invocation and response.
Designing linearizable implementations is a simple task
due to Herlihy's Universal Construction~\cite{H91},
although the resulting solutions typically do not scale well in practice.
In contrast, designing linearizable \emph{and} scalable implementations is a challenging task,
as it requires 
{the use of fine-grained locking or the absence of locks at all},
in order to exploit the parallelism in concurrent systems,
which typically derives in a large number of scenarios and subtle corner cases 
that need to be considered in correctness proofs~\cite{HS08,MS04,R13}. 

The importance of linearizability, and more broadly of correct concurrent software, 
naturally calls for formal verification techniques.
Over the past years, a variety of techniques for verifying linearizability 
have been developed, using different approaches and providing different levels of guarantees.
See for example survey~\cite{DD15}.
Despite of all efforts, verifying correctness of linearizable implementations
remains difficult. Model checking is feasible only for small cases 
(i.e. for a bounded number of processes, and/or invocations to operations),
and finding linearization points, simulations and invariants is hard, sometimes 
{requiring deep understanding of the verified implementation}.
The problem has been also studied from a theoretical
perspective. It is known that 
deciding whether an implementation is linearizable might be EXPSPACE-complete or even undecidable~\cite{BEEH13},
while deciding if a given finite execution is linearizable might be NP-complete~\cite{GK97,P79}.

\subsection{Distributed runtime verification and its challenges}

Runtime verification is a dynamic, lightweight, yet rigorous, formal method that complements
static verification techniques with a more practical approach. 
It only seeks to verify that the current execution of a system is correct,
and maybe prevent an incorrect action or enforce a correct behavior otherwise.
The system under inspection can be of any type, from hardware to software, centralized or distributed.
We refer the reader to~\cite{BF18, FHR13, HG05, LS09} for a detailed exposition of the field.

Broadly speaking, in runtime verification, two, non necessarily disjoint tasks need to be accomplished~\cite{BFRT16}:
(1) the design of a \emph{communication interface} that detects the current execution 
of the \emph{underlying system} under inspection, and (2) the design of a \emph{monitoring system} that verifies 
whether the detected execution is correct with respect to some correctness criteria.
A big source of difficulty is that 
the underlying system and the communication interface are typically \emph{decoupled}.
Namely, the underlying system is designed, implemented and deployed
without considering that in the future a runtime verification mechanism might be integrated to it,
hence it might not export enough data of the current execution from
which a communication interface and a monitoring system can be built later on.
The situation gets more difficult when the underlying system is \emph{distributed} 
as no process of the system ``knows'' what the current execution is, 
each process has only a \emph{partial view} of what the execution could be. 
The problem is even worse if we seek for a \emph{distributed, asynchronous and fault-tolerant}
communication interface, as the processes might not even 
have the ability to agree on the partial view of a process of the underlying system
(e.g.~\cite{FLP85}).
In such scenario, there are several computational entities that exchange information,
subject to delays and failures, yet they have to make consistent decisions.
Namely, we have a distributed computing problem.

Designing distributed runtime verification algorithms that are asynchronous and fault-tolerant
is a challenging problem (see~\cite{BFRT16, E-HF18, F21, FPS18, LFKV18, SSABBCFFK19}).
In fact, there are only runtime verification solutions (for a number of properties) 
that are \emph{failure-free} and \emph{synchronous} (e.g.~\cite{AFIMP20, SHKKNPPW12}),
fault-tolerant with \emph{timing assumptions} 
(e.g.~\cite{BKZ15, BGKS20, BF16, BFRR22, E-HF20, FRT13, FCF14, FRT20, GXJLSBH22, KB18, RF22}),
or asynchronous \emph{failure-free} (e.g.~\cite{CGNM13, FF08, FFY08, LR13, SS14, SVAR04}).
That is, the known distributed runtime verification algorithms are \emph{not fully asynchronous and fault-tolerant}.
As far as we know, runtime verification of linearizability has only been studied in~\cite{ETQ05, ET06},
in asynchronous failure-free models with centralized monitoring systems.

\subsection{Asynchronous wait-free runtime verification of linearizability}

We are interested in runtime verification of linearizability, where the underlying system 
\m{A} is an asynchronous concurrent\footnote{{Through the paper, we interchangeably use the words `distributed' and `concurrent'
to denote situations where several processes communicate each other in order to
collectively solve a problem. In the context of shared memory algorithms, we always use `concurrent'.}} 
implementation of an object,
and the communication interface and monitoring system are consolidated in an asynchronous wait-free
shared memory concurrent algorithm \m{V}.
\emph{Wait-free}~\cite{H91} means that the algorithm tolerates any number of crash failures.
Moreover, we focus on the case where the underlying system \m{A} is a \emph{black-box},
i.e., we do not have access to its specification/pseudocode, hence the only way 
to obtain information of the current execution is by analyzing the sequence of 
invocations and responses each process obtains from~it.

 \begin{figure}[ht]
\begin{center}
\includegraphics[scale=0.75]{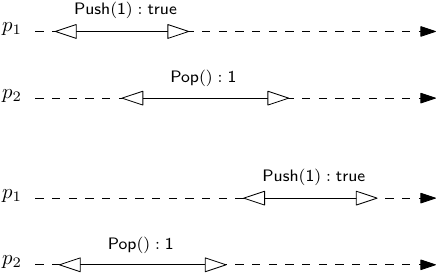}
\vspace{-0.1cm}
\caption{Two executions of a stack where the two processes have the same
 partial views, i.e. sequence of invocations and responses;
 while the execution at the top is linearizable, the execution at the bottom is not.
 Real-time, unaccessible to the processes, is what ultimately defines the executions.}
\label{fig-example-partial-views}
\end{center}
\end{figure}

Why is this setting interesting? 
While arguably \m{A} and \m{V} being asynchronous and \m{A} being a black-box
model the challenging scenario described above, 
where the system under inspection and the communication interface are decoupled, 
\m{V}~being asynchronous and wait-free guarantees
that {the progress and timing properties of \m{A} are preserved}. 
For example, if \m{A} is non-blocking (e.g., wait-free) but \m{V} is blocking
(maybe because it lock-based), the system that results of integrating \m{A} and \m{V} 
 will be blocking, i.e. non fault-tolerant, 
hence {weakening} \m{A}'s progress property.
Similarly, \m{A} might be designed as an asynchronous implementation for sake efficiency, 
hence if \m{V} is synchronous or semi-synchronous, 
\m{A}'s timing property is {weakened}.
Indeed, a target in runtime verification is to design communication interfaces and monitoring systems
that are ``as less intrusive as possible''; ideally they should not interfere with the 
behavior of the underlying system (e.g.~\cite{BBF15, BF18, DGHLSSW17, SEF23}).

{In presence of asynchrony and failures}, is it possible to capture the actual execution of \m{A} 
in order to verify linearizability? 
The answer is clearly no. After the seminal work of Lamport~\cite{L78}, we know that it is \emph{impossible} to
determine the order of \emph{non-causally related} events in fully asynchronous distributed systems,
and invocations to and responses from \m{A} 
are local events, which are non-causally related.
The partial view of a process in \m{A}'s actual execution is just the sequence with its invocation and responses,
and for the processes is just impossible to access the real-time order of these local events,
which ultimately defines \m{A}'s execution (see Figure~\ref{fig-example-partial-views}).
This reasoning {seems to imply} that very little should be possible when considering linearizability, as in the end
this correctness condition totally depends on real-time order of locals events.
However, there are ways to runtime verify linearizability, as we show here.

\subsection{Results}

We propose an interactive model {for the theoretical study of} the problem of distributed 
runtime verification of correctness conditions.
In the model, a concurrent implementation~\m{A} 
interacts with a \emph{verifier}~\m{V},
which is required to invoke operations of \m{A},
receive the corresponding responses, and somehow compute relevant information of the current execution of \m{A},
in order to decide whether it is correct or not;
the interaction between \m{A} and \m{V} is infinite.
The verifier \m{V} \emph{runtime verifies} linearizability, if \m{V} is \emph{sound},
{i.e., if \m{A}'s current execution is linearizable, \m{V} does not report $\error$},
and \emph{complete}, 
{namely, if \m{A}'s current execution is not linearizable, \m{V} reports $\error$}.

Using a simple indistinguishability argument, we show that 
it is impossible to runtime verify linearizability for some common objects such as 
queues, stacks, sets, priority queues, counters and even the consensus problem,
regardless of the \emph{consensus number}~\cite{H91} of the base objects used in~\m{V}.
{Thus, differently from the case of linearizability for which consensus suffices for obtaining 
linearizable wait-free implementations for any object~\cite{H91}, wait-free runtime verification of
linearizability of some objects is impossible even if consensus is available.}

Then, perhaps somewhat surprisingly, we show that a variant of the problem can be solved,
and without the need of consensus,
if linearizability is verified \emph{indirectly}. 
Namely, we identify a class of concurrent implementations, called {\emph{Distributed Runtime Verifiable} (\RV)},
such that a variant of the {distributed runtime verification problem} can be solved with respect to \RV.
In the variant, a verifier \m{V} is required to be complete and \emph{predictively sound}, 
{which intuitively means that \m{V} is allowed to report \error when \m{A}'s current execution is linearizable (i.e., a \emph{false negative}),
as long as it ``predicts'' that a \m{A} is \emph{not linearizable},
namely, \m{V} possesses an execution of \m{A} that is not linearizable 
(hence that non-linearizable execution is necessarily different from the current one).}

Intuitively, any implementation in the class \RV produces a ``sketch'' of its current execution, additionally to operation outputs.
It turns out that such sketches are good enough to predictively verify linearizability, 
for the class~\RV, and using only read/write base objects, 
which have consensus number one~\cite{H91} (hence
incapable of solving consensus among two or more processes).
Linearizability can then be indirectly predictively verified because \emph{any}
concurrent implementation~\m{A} can be transformed
into an implementation $\m{A}^*$ in \RV, using only read/write base objects, 
such that~\m{A} and~$\m{A}^*$ have the same progress properties, and 
\m{A} is linearizable 
if and only if $\m{A}^*$ is linearizable, both with respect to the same sequential object. 
Therefore, although verifying linearizability of~$\m{A}$ is impossible,
it is possible to derive an implementation~$\m{A}^*$ from~$\m{A}$ that implements the same object and
linearizability of $\m{A}^*$ is predictively verifiable.

{As a by-product, we obtain a methodology to derive implementations
that verify linearizability of any of its response at runtime, or report $\error$ if they cannot produce
linearizable responses any longer. Moreover, the implementations are able to provide 
an execution that \emph{certifies} whether the responses so far are linearizable or not.
We call such implementations \emph{self-enforced linearizable}.}
The methodology takes \emph{any} implementation~\m{A} and 
produces an implementation \m{B} using $\m{A}$ as a black-box (to produce operation responses) 
and read/write objects (to runtime verify \m{A}'s responses) 
such that \m{A} and \m{B} have the same progress properties and
\m{A} is linearizable if and only if \m{B} is self-enforced linearizable.
Thus, in a concurrent system, one can use the derived self-enforced implementation~\m{B} 
instead of~\m{A} with the guarantee that all non-$\error$ responses of~\m{B}
are linearizable, as all of then are runtime verified by~\m{B} itself.
We are not aware of prior concurrent implementations in the literature with such properties.
Due to their capability to certify linearizability of their responses,
we believe self-enforced linearizable implementations allow the design of concurrent systems 
in a modular manner (due to modularity of linearizability~\cite{HW90, SHP21}) with accountable and forensic guarantees. 
Remarkably, the proposed methodology is simple and generic, hence 
prone to being program synthesized 
{(i.e., automatically constructing~\m{B} from \m{A} with \m{B} guaranteed to have the aforementioned properties)}, 
a property of interests in runtime verification.

All results hold for a correctness condition \GenLin defined here, 
which includes linearizability,
as well as {set-linearizability}~\cite{N94} and
{interval-linearizability}~\cite{CRR18}, both generalizations of linearizability 
for concurrent objects with no sequential specifications~\cite{CRR23}.

\subsection{Structure of the paper}

Once Section~\ref{sec-model} states the model of computation, the interactive model for distributed runtime verification
and the problem of distributed runtime verification are introduced in Section~\ref{sec-model-verification}.
Then, Section~\ref{sec-linearizability} recalls the definition of linearizability and 
Section~\ref{sec-impossibility} shows the impossibility for runtime verification of linearizability for some common objects.
A high-level perspective of how the impossibility result is evaded and the definition
of the predictive version of the problem appear in Section~\ref{sec-overview}.
Sections~\ref{sec-class-DRV} and~\ref{sec-DRV-verifiable} implement 
the high-level strategy described in Section~\ref{sec-overview}.
Section~\ref{sec-extensions} presents some extension of our results.
Section~\ref{sec-related-work} discusses related work, explaining differences with our results,
and Section~\ref{sec-final} concludes the paper with a final discussion.

\section{Model of Computation}
\label{sec-model}

We consider a standard concurrent shared memory system (e.g.~\cite{HS08, HW90}) with
\(n \geq 2\) \emph{asynchronous} processes, \(p_1, \hdots, p_n\),
which may \emph{crash} at any time during an execution. 
All but one process can crash in any execution.
The \emph{index} of process \(p_i\) is \(i\).
Processes
communicate with each other by invoking \emph{atomic} operations of
shared \emph{base objects} that reside in the shared memory: either simple \R/\W operations, or more
complex and powerful \RMW operations, such as \FAI or \CAS. 
In algorithm descriptions, shared base objects are denoted with uppercase letters;
local variables used by a process for performing its local computations are 
denoted with lowercase letter subscripted with the index of the process.

{We do not provide detailed specifications of \RMW base operations
because (1) the proof of the impossibility result in Theorem~\ref{theo-impossibility}
does not need them, and (2) our algorithms use only standard \R/\W operations.
We refer the reader to standard textbooks such as~\cite{HS08} for definitions of those operations.
The point of mentioning \RMW operations in this section is that Theorem~\ref{theo-impossibility} holds
even if any of those operations are used in a verifier (the definition of verifier appears in Section~\ref{sec-model-verification}).}

For ease of exposition, it is assumed that base objects are of unbounded size.
Section~\ref{sec-final} explains how this unrealistic assumption can be removed from the proposed algorithms.
We consider the possibility that processes have \emph{perfectly synchronized local clocks}. 
Each process can read its local clock in a local computation step, information that then can be written 
in the shared memory. We stress that the time that elapses between reading a local clock and
writing it in the shared memory is unpredictable, as the system is asynchronous.
This assumption is not relevant in our algorithms, however it makes our impossibility result in Theorem~\ref{theo-impossibility} stronger.

An \emph{implementation} of a concurrent object $\m{O}$ (e.g. a queue or a stack), specified in some way 
(more details in Section~\ref{sec-linearizability}),
is a distributed algorithm $\mathcal A$ consisting of $n$ local state machines 
$A_1, \hdots, A_n$, some of them possibly non-deterministic.  
Local machine $A_i$ specifies which operations on base objects and local computations 
$p_i$ executes in order to return a response when it invokes a high-level
operation of $\m{O}$ (sometimes simply called operation).  
Each of these base objects operations and local computations is a \emph{step}.
Invocations and responses are local computations as well but 
we do not refer to them as steps.

An \emph{execution} of $\mathcal A$ is a possibly infinite sequence of
steps, plus invocations
and responses to high-level operations of the concurrent object $\m{O}$,
with the following \emph{well-formedness} properties:
\begin{enumerate}
\item Each process is sequential, namely, it first invokes a high-level operation, and only when
  it has a corresponding response, it can invoke another high-level
  operation.

\item A process takes steps only between an invocation and a response.

\item For any invocation of  an operation $\op_i$ {by} a process $p_i$,
denoted $inv_i({\op_i})$, the steps of $p_i$ between that
  invocation and its corresponding response (if there is one), denoted
  $res_i({\op_i})$, are steps specified by $A_i$ when $p_i$ invokes~$\op_i$.
\end{enumerate}

It is assumed that after a process completes an operation,
it non-deterministically picks the operation it executes next.
{This assumption is to guarantee that correctness is proven considering
all possible combinations of invocations to high-level operations of an implementation.}

For simplicity, and without loss of generality, we assume that every 
concurrent object provides a single high-level operation, called $\sf Apply$, that receives
as input $\op$, a description of  the actual operation that is invoked, which includes 
the inputs to the actual operation. 
We also assume, again without loss of generality,
 that $\sf Apply$ is invoked with a given input $\op$ only once
(a fictitious input value to the actual operation can be added such that all inputs are different).

Typically, invocation and responses of an implementation \m{A} also include \m{A} 
as part of their description. We however drop that information from invocations
and responses because it facilitates our discussion, although some 
ambiguity is introduced.

An implementation \m{A} can use other implementation \m{B} in order
to produce responses, namely, the processes can invoke 
high-level operation of \m{B}. Hence, when a process invokes an
operation of~\m{B}, it continues executing the steps specified by \m{A}
only after it receives the corresponding response from \m{B}.
Consider any execution $E$ of \m{A}.
We let $E|\m{B}$ denote the sequence of steps, invocations and responses of \m{B} in $E$.
Then, $E|\m{B}$ is a well-formed execution of \m{B}.
In what follows, unless stated otherwise, when we talk about operations in
$E$, we mean only the operations of \m{A}, excluding the nested calls to operations of \m{B}.

A high-level operation in an execution is \emph{complete} if
both its invocation and response appear in the execution.
An operation is \emph{pending}
if only its invocation appears in the execution.  
A process is \emph{correct} in an infinite execution of an implementation if it takes
infinitely many steps. When considering infinite executions, we focus on those that are \emph{fair}:
for every correct process and every~$K \geq 1$, there is a finite prefix
with $K$ steps of that process.
An implementation \m{A} is \emph{wait-free} if in
every infinite execution, every correct process completes infinitely
many operations~\cite{H91}.  An implementation \m{A} is \emph{lock-free}
if in every infinite execution, infinitely many operations
are complete~\cite{HW90}.  Thus, a wait-free implementation is
lock-free but not necessarily vice versa.
We consider only implementations that are at least lock-free.
The notions of wait-freedom and lock-freedom naturally extend to 
specific operations or fragments of pseudocode.

The \emph{step complexity} of an implementation is the maximum number
of base operations a process needs to take to produce a response.

Sometimes it will be convenient to think an implementation \m{A} as a \emph{black-box} 
whose specification cannot be accessed, and hence the only information that can
be obtained from it are the executions it produces without steps.
We call such executions without steps \emph{histories}, namely, sequences
of invocations and responses satisfying the first two well-formedness properties stated above.
As in~\cite{HW90}, we define an \emph{abstract implementation} as a set of well-formed histories.
By abuse of notation, for any execution $E$ of an implementation \m{A}, 
we let $E$ itself denote the history obtained from $E$ 
(i.e. the sequence obtained by removing from $E$ all its steps), 
and let \m{A} denote itself the 
abstract implementation with all histories of~\m{A} 
(i.e., the set with all histories obtained from the executions of~\m{A}).
This abuse of notation will facilitate the discussion, at the cost of introducing some ambiguity.

The \emph{consensus number}~\cite{H91} of a shared object $\m{O}$ is the maximum
number of processes that can solve the \emph{consensus}
problem, using any number of instances of $\m{O}$ in addition to any
number of \R/\W base objects.  Consensus numbers induce the
\emph{consensus hierarchy} where objects are classified according
their consensus numbers.  The simple \R/\W operations stand at the
bottom of the hierarchy, with consensus number one and the lowest coordination power.
At the top of the hierarchy we find operations with infinite consensus
number, like \CAS, that provide the maximum possible coordination power.

\section{An Interactive Model for Distributed Runtime Verification}
\label{sec-model-verification}

Let us fix a concurrent object $\m{O}$, specified in some way. 
Let $\m{A}$ be a lock-free implementation of $\m{O}$.
Intuitively, a \emph{correctness condition} is a mechanism to separate the correct
implementations of $\m{O}$ from the incorrect ones. Basically, it is a predicate ${\sf P} _\m{O}$ that \emph{all}
finite executions of $\m{A}$ need to satisfy for $\m{A}$ being declared correct, 
with respect to ${\sf P} _\m{O}$. It is known that if ${\sf P} _\m{O}$ is linealizability,
deciding whether \m{A} is linearizable might be EXPSPACE-complete or even undecidable~\cite{BEEH13},
depending on the object~$\m{O}$.
Deciding if a given finite history is linearizable is decidable, but it might be NP-complete~\cite{GK97,P79}, 
although for some objects this question can be decided in polynomial time~\cite{BEEH15, EE18}.
From now on, we will assume that each process can locally test 
if a given finite history satisfies ${\sf P} _\m{O}$.

\begin{figure}[ht]
\centering{ \fbox{
\begin{minipage}[t]{150mm}
\scriptsize
\renewcommand{\baselinestretch}{2.5} \resetline
\begin{tabbing}
aaaa\=aaa\=aaa\=aaa\=aaa\=aaa\=aaa\=\kill 

{\bf Shared Variables:}\\

$~~$  Shared memory $M$\\ \\

{\bf Operation}  $\Verify(\m{A})$ {\bf is} \\

\line{L01} \> $set_i \leftarrow \emptyset$\\

\line{L02} \> {\bf while} {\sf true} {\bf do}\\

\line{L03} \>\> $\op_i \leftarrow$ non-deterministically chosen high-level operation that is not in $set_i$\\

\line{L04} \>\> $set_i \leftarrow set_i \cup \{\op_i\}$\\

\line{L05} \>\> Encode in $M$ the invocation to ${\Apply}(\op_i)$ of $\m{A}$\\

\line{L06} \>\> Invoke operation ${\Apply}(\op_i)$ of \m{A} \%\% Local even of process $p_i$\\

\line{L07} \>\> $resp_i \leftarrow$ response from operation ${\Apply}(\op_i)$ of \m{A} \%\% Local even of process $p_i$\\

\line{L08} \>\> Encode the response $resp_i$ in $M$\\

\line{L09} \>\> $exec_i \leftarrow$ description of the current execution of $\m{A}$ in $M$\\

\line{L10} \>\> {\bf if} $\neg \, {\sf P}_O(exec_i)$ {\bf then} \\

\line{L11} \>\> \> {\bf report} $(\error, exec_i)$\\

\line{L12} \>\> {\bf end if} \\

\line{L13} \> {\bf end while}\\

{\bf end} \Verify

\end{tabbing}
\end{minipage}
  }
\caption{\small Generic structure of a verifier $\m{V}_\m{O}$ for correctness condition ${\sf P}_\m{O}$ (code of process $p_i$).}
\label{fig-verifier}
}
\end{figure}

Let us suppose the existence of a \emph{client} $\m{C}$ (i.e. a concurrent algorithm) that solves some distributed problem 
using~\m{A}, {assuming that \m{A} is correct with respect to ${\sf P} _\m{O}$}.
We would like to design an \emph{intermediate layer} $\m{V}_\m{O}$ between \m{C} and \m{A} that, 
from time to time, verifies that the \emph{current} execution of \m{A} is correct, namely, it satisfies the predicate ${\sf P} _\m{O}$,
and reports $\error$ otherwise
(basically $\m{V}_\m{O}$ consolidates a communication interface and a monitoring system in a single entity). 
If \m{A} is indeed correct, then
we would like the client \m{C} not to be able to distinguish whether it is interacting 
with~\m{A} or~$\m{V}_\m{O}$, and hence we require $\m{V}_\m{O}$ to be asynchronous and wait-free
so that the properties of~\m{A} are preserved (e.g., its progress properties).

We model this layer as a concurrent algorithm $\m{V}_\m{O}$, 
called \emph{verifier},
that interacts with $\m{A}$. The generic structure of the interaction appears in Figure~\ref{fig-verifier},
where $\m{A}$ is a \emph{black-box}, and hence the only way
$\m{V}_\m{O}$ can obtain information of $\m{A}$ is by invoking high-level operations of it;
namely, the processes \emph{only} invoke operations of $\m{A}$ and \emph{only} receive responses.
Thus, $\m{V}_\m{O}$ interacts with an \emph{abstract implementation} $\m{A}$ (as defined in Section~\ref{sec-model}),
from which $\m{V}_\m{O}$ receives one of its histories in every execution. 

During the interaction, each process invokes a series 
of non-deterministically chosen high-level operations of $\m{A}$ 
in order to consider all possible combination of operations of \m{A} that any client~\m{C} might invoke. 
For simplicity, every process tests if the 
history so far satisfies ${\sf P}_\m{O}$ after each of its high-level operations of \m{A} (Line~\ref{L10}).
If the predicate is not satisfied, the process reports \error together with a \emph{witness} to \m{A},
i.e. a history of \m{A} that does not satisfy ${\sf P}_\m{O}$;
in any case, the interaction continues.
(In a practical setting the interaction would stop, and reporting \error and the witness would be returned to the client~\m{C}; 
for simplicity, in our model the interaction continues.)
Naturally, the processes in $\m{V}_\m{O}$ need to exchange information in order to
obtain a description of the current history of $\m{A}$.
Thus, each process might store information in the shared memory
before and after each of its invocations to and responses from~$\m{A}$ (Lines~\ref{L05} and~\ref{L08}).\footnote{The communication 
in Lines~\ref{L05} and~\ref{L08} ultimately creates causal relations~\cite{L78} between invocations to and responses from~$\m{A}$,
whose aim is to compute the current execution of \m{A}.}

Since $\m{A}$ is a black-box in the generic verifier in Figure~\ref{fig-verifier}, $\m{V}_\m{O}$ cannot be designed for a specific
implementation $\m{A}$, it must work for \emph{any} possibly abstract concurrent implementation
(even if it is correct with respect to an object $\m{O}' \neq \m{O}$).
This requirement is modeled as \emph{conceptually} $\m{V}_\m{O}$ taking~\m{A} as its input to the computation.

We restrict our attention to verifiers where the block codes in
Lines~\ref{L03}--\ref{L05} and Lines~\ref{L08}--\ref{L12} are wait-free.
We say that such verifiers are \emph{wait-free}.
The \emph{step complexity} of a wait-free verifier is the maximum number of base operations
a process needs to take in order to complete one iteration of the while loop,
discarding steps in $\m{A}$.

In the definition of the distributed runtime verification problem below, for simplicity, it is assumed that \emph{no process crashes},
and hence all executions of a verifier are infinite, since we are assuming that~$\m{A}$ is lock-free or wait-free.
The assumption makes the problem easier to state. The possible crashes that can occur in the system
are modeled by the fact (1) we focus on wait-free verifiers, (2) the system is asynchronous and 
(3) correctness is tested at finite prefixes of a given infinite execution.
The soundness and completeness properties that specify the problem have been considered in the past 
(see for example~\cite{E-HF18, MB15, NFBB17}); 
the definition basically adapts them to fit in our interactive setting.

\begin{definition}[Distributed Runtime Verification]
Let $\m{O}$ be a concurrent object specified in some way, and consider a correctness condition
${\sf P}_\m{O}$ for $\m{O}$. We say that a \emph{wait-free} verifier $\m{V}_\m{O}$ \emph{distributed runtime verifies} ${\sf P}_\m{O}$
if the following two requirements are fulfilled in every infinite execution $E$ of $\m{V}_\m{O}$
with an arbitrary input (abstract) implementation~\m{A}:
\begin{enumerate}
\item Soundness: If for every finite prefix $E'$ of $E$, $E'|\m{A}$ satisfies ${\sf P}_\m{O}$, 
then no process reports \error.

\item Completeness. If $E$ has a finite prefix $E'$ such that $E'|\m{A}$ does not satisfy~${\sf P}_\m{O}$, 
then at least one process reports \error together with a witness to \m{A}.
\end{enumerate}

We say that ${\sf P}_\m{O}$ is \emph{distributed runtime verifiable} if there is a wait-free verifier 
that distributed runtime verifies ${\sf P}_\m{O}$.
\end{definition}

The previous definition is flexible, it is not difficult to modify it 
to cover the cases where $\m{A}$ is {blocking}~\cite{HS08}
(i.e. it internally uses {locks}) or {obstruction-free}~\cite{HS08} 
(namely, progress is guaranteed only when a process runs solo), 
or correctness conditions for one-shot distributed problems such as {tasks}~\cite{HKR13},
where each process invokes one high-level operation.
The main difference is that in these cases, the interaction might be only finite.

Below, for sake of compactness, sometime we will simply say verify/verifiable instead of distributed runtime verify/verifiable. 

\section{Linearizability}
\label{sec-linearizability}

\begin{figure}[ht]
\begin{center}
\includegraphics[scale=0.65]{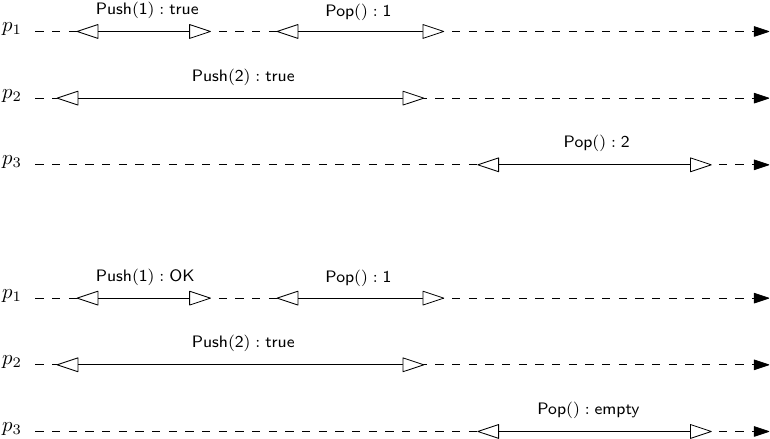}
\vspace{-0.1cm}
\caption{Two 3-process histories of a stack implementations are depicted. 
Time goes from left to right, and operations are denoted with a double-ended arrows.
For clarity, each operation $\Apply(\op)$ is simply denoted~$\op$.
The execution at the top is linearizable with respect to the usual sequential specification of a stack; a linearization 
of it is: $\sf \langle Push(2): true \rangle  \langle Push(1): true \rangle  \langle Pop(): 1 \rangle  \langle Pop(): 2 \rangle$.
The execution at the bottom is not linearizable because the stack cannot be empty when the operation
$\sf \langle Pop(): empty \rangle$ starts.}
\label{fig-examples-lin-seq-cons}
\end{center}
\end{figure}

\emph{Linearizability}~\cite{HW90} is the de facto standard 
correctness condition for concurrent implementations of objects defined with
sequential specifications. It extends  the concept of \emph{atomicity}~\cite{L86a, L86b}
to any sequential object.
Intuitively, an execution of an implementation is linearizable if its operations can be ordered sequentially,
without reordering non-overlapping operations, so that their responses
satisfy the sequential specification of the implemented object.
Figure~\ref{fig-examples-lin-seq-cons} depicts examples 
of linearizable and non-linearizable histories of a stack implementation.

\begin{definition}[Sequential Specifications]
A \emph{sequential specification} of a concurrent object~$\m{O}$ is a
state machine specified through a (possibly partial) transition function $\delta$. 
Given a state $q$ and an invocation $inv_i({\op_i})$ of process $p_i$,
$\delta(q, inv_i({\op_i}))$ returns the tuple $(q', res_i({\op_i}))$
indicating that the machine moves to state $q'$ and the response to
$\op_i$ is $res_i({\op_i}$).
{If the machine is \emph{non-deterministic}, $\delta(q, inv_i({\op_i}))$ is 
a set of tuples from which one of them is non-deterministically chosen
to respond to $inv_i({\op_i})$ and move the machine to the next state.}
  The sequences of invocation-response tuples, $\langle inv_i({\op_i}): res_i({\op_i}) \rangle$, 
  produced by the state machine are
  its \emph{sequential histories}.  
 \end{definition}

 For sake of clarity, a tuple
  $\langle inv_i({\op_i}): res_i({\op_i}) \rangle$ is simply denoted
  $\op_i$. Also, subscripts of invocations and responses are omitted.

A history $E'$ is an \emph{extension} of a finite history $E$, 
if $E'$ can be obtained from $E$  by appending zero or more responses 
for some of $E$'s pending operations.  
  
For any history $E$ and any process $p_i$, $E|p_i$
denotes the sequence of invocations and responses of~$p_i$ in $E$.
Two histories $E$ and $F$ are \emph{equivalent} 
if $E|{p_i} = F|{p_i}$, for every process~$p_i$.
  
For any finite history $E$ of an implementation \m{A}, $comp(E)$ denotes the history
obtained by removing from $E$ all invocations of pending operations;
note that $comp(E)$ is well-formed.
To formalize linearizability, we define a partial order $<_E$ on the
\emph{complete} operations of any history~$E$: ${\op} <_E {\op}'$ if
and only if $res({\op})$ precedes $inv({\op}')$ in $E$.  Two
complete operations are \emph{concurrent} if they are incomparable by
$<_E$.  The history is \emph{sequential} if $<_E$ is a total order.

We consider the definition of linearizability in~\cite{SHP21}, 
which is a slight variant of the original definition~\cite{HW90} that fixes some corner cases.

\begin{definition}[Linearizability]
Let $\m{O}$ be any concurrent object. 
A finite history $E$ is \emph{linearizable} with respect to $\m{O}$ 
if there is an extension $E'$ of $E$ and a sequential history $S$ of $\m{O}$ such that 
\begin{enumerate}
\item $comp(E')$ and $S$ are equivalent and 
\item $<_{comp(E')} \, \subseteq \, <_S$.
\end{enumerate}
The sequential history $S$ is said to be a \emph{linearization} of $E$.
We say that an implementation $\mathcal A$ is \emph{linearizable} with respect to~$\m{O}$, 
if each of its finite histories is linearizable with respect to $\m{O}$.
\end{definition}

\section{Linearizability is not Runtime Verifiable}
\label{sec-impossibility}

This section shows that for some common objects such 
queues, stacks, priority queues, counters,
and even the fundamental consensus problem,  linearizability is not distributed runtime verifiable, 
regardless of the consensus number of the base objects 
the processes use in a verifier.
The following simple impossibility proof captures informal arguments that have been used in the past (e.g.~\cite{E-HF18, GMB20})
to argue that distributed runtime verification of some correctness conditions in asynchronous distributed systems is impossible.

\begin{theorem}[Impossibility for Distributed Runtime Verification of Linearizability]
\label{theo-impossibility}
 For queues, stacks, sets, priority queues, counters
and the consensus problem (defined as sequential objects) linearizability is not 
distributed runtime verifiable, 
{regardless of the consensus number of base objects used in a verifier.}
\end{theorem}

\begin{proof}
We focus on the case of the queue as all other cases are very similar.
By contradiction, suppose that there is a wait-free verifier $\m{V}_{\sf queue}$ that
verifies linearizability for queues. 
Consider the following \emph{non-linearizable} implementation~\m{A}:
every \Enq operation returns $\sf true$, and every \Deq operation returns $\sf empty$,
for every process $p_i \neq p_2$, and for $p_2$, it returns $1$ in its first operation,
and returns  $\sf empty$ in every subsequent operation.

\begin{figure}[ht]
\begin{center}
\includegraphics[scale=0.75]{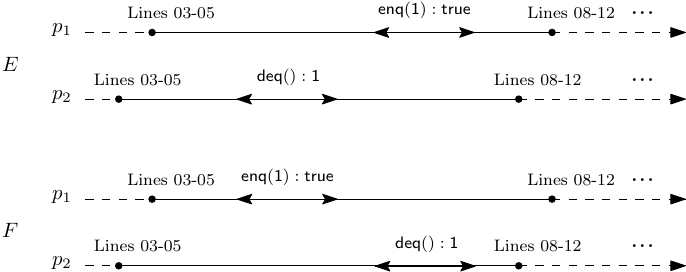}
\caption{\small Executions $E$ and $F$ in the impossibility proof for $n = 2$.}
\label{fig-example-proof}
\end{center}
\end{figure}

We will exhibit two executions $E$ and $F$ of $\m{V}_{\sf queue}$ with input \m{A} and argue that 
$\m{V}_{\sf queue}$ cannot simultaneously satisfy the soundness and completeness requirements of 
the distributed runtime verification problem.
We use the generic structure in Figure~\ref{fig-verifier} to describe the executions.

Execution $E$ is the next (see Figure~\ref{fig-example-proof}):
\begin{enumerate}
\item In its first iteration of the while loop, process $p_2$ picks $op_2 = \Deq()$ in Line~\ref{L03},
executes the local step in Line~\ref{L04} and
the base operations corresponding to Line~\ref{L05}.

\item In its first iteration of the while loop, process $p_1$ picks $op_1 = \Enq(1)$ in Line~\ref{L03},
executes the local step in Line~\ref{L04} and
the base operations corresponding to Line~\ref{L05}.

\item Process $p_2$ executes Lines~\ref{L06} and~\ref{L07} of its first iteration of the while loop.
Thus it obtains response $resp_2 = 1$ for its high-level operation $op_2 = \Deq()$.

\item Process $p_1$ executes Lines~\ref{L06} and~\ref{L07} of its first iterations of the while loop.
Thus it obtains response $resp_1 = {\sf true}$ for its high-level operation $op_1 = \Enq(1)$.

\item Process $p_2$ executes Lines~\ref{L08} to~\ref{L12} of its first iteration of the while loop.

\item Process $p_1$ executes Lines~\ref{L08} to~\ref{L12} of its first iteration of the while loop.

\item For each $k = 1, 2, \dots, \infty$ (in this order), 
$p_{(k\mod n)+1}$ executes a whole iteration of the while loop where it picks $op_{(k\mod n)+1} = \Deq()$ in
Line~\ref{L03} (and hence the response in Line~\ref{L07} is $resp_{(k\mod n)+1} = {\sf empty}$).
\end{enumerate}

Execution $F$ is similarly constructed, with the exception that the steps 3 and 4 of the
previous construction appear in the opposite order (see Figure~\ref{fig-example-proof}).
In other words, in $E$, the first high-level operations of $p_2$ is executed first and
then the first high-level $op_1$ is executed, whereas in $F$, the high-level operations
are executed in the opposite order. 

The history of \m{A} obtained from every finite prefix of $F$
is linearizable; in contrast, the history of~\m{A} obtained from every finite prefix of $E$ containing 
at least the first operation of $p_2$, is not.

As the only difference between $E$ and $F$ is the order of occurrence of local events of $p_1$ and $p_2$,
and this order is not accessible to $p_1$ and $p_2$,
every process $p_i$ transitions through the same sequence of local states in both
executions $E$ and $F$.
This means that the executions are \emph{indistinguishable} to all processes,
and hence in both executions they make the same sequence of decisions in Lines~\ref{L08} to~\ref{L10}.
Thus, if no process reports \error in $E$ and $F$, then $\m{V}_{\sf queue}$ does not fulfills completeness due to~$E$,
and if at least one process reports \error in $E$ and $F$, then $\m{V}_{\sf queue}$ does not fulfills soundness due to $F$.
Therefore, $\m{V}_{\sf queue}$ cannot exist.

For the other objects, the argument is nearly the same. For example, for the case of the stack, 
\Pop is replaced with \Deq, and \Push is replaced with \Enq. For the case of the consensus, 
we define an object with a single {\sf Decide} operation that can be invoked several times,
and the first operations among all processes sets its input as the decision of the consensus.
\end{proof}

The previous proof can easily be extended to 
variants of linearizability 
that have been used to specify relaxed versions of sequential objects.
Examples of such variants are {quasi-linearizabilty}~\cite{AKY10},
{$k$-stuttering}~\cite{HKPSS13}, {set-linearizability}~\cite{N94}, {interval-linearizability}~\cite{CRR18}
and {intermediate value linearizability}~\cite{RK20}.
The reason is that all these relaxations include the sequential executions with
the ``exact'' sequential behavior of the object that is relaxed, which suffices for the previous proof to~hold.

\section{Evading the Impossibility Result: The Strategy at a High-Level}
\label{sec-overview}

The rest of the paper is devoted to show that it is possible to  
circumvent the impossibility in Theorem~\ref{theo-impossibility}.
Roughly speaking, it will argue that, for any sequential object $\m{O}$, 
linearizability of any implementation \m{A} of \m{O}
can be \emph{indirectly} verified through
a class of implementations that we call {\emph{Distributed Runtime Verifiable} (\RV)}.
Perhaps somewhat surprisingly, {the class \RV can be verified with respect to a \emph{predictive} 
variant of the distributed runtime verification problem, in which, generally speaking,
a verifier \m{V} is allowed to report a \emph{false negative} as long as \m{V}
``predicts'' that \m{A} is not linearizable, namely, \m{V} has evidence of a non-linearizable execution of \m{A}.}
This section gives a high-level perspective of the strategy followed in the next sections,
and introduces the predictive variant of the distributed runtime verification problem.

\begin{figure}[ht]
\begin{center}
\includegraphics[scale=0.65]{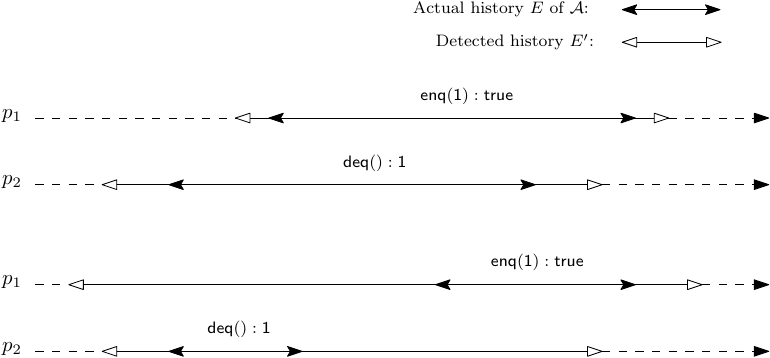}
\vspace{-0.1cm}
\caption{Due to asynchrony, processes in a verifier might detect histories where operations of $\m{A}$ are ``stretched''.
This phenomenon is exemplified with two examples of a queue implementation 
where an operation $\Apply(\op)$ is simply denoted $\op$.
At the top, both histories, the actual one $E$ and the detected one $E'$, are linearizable.
At the bottom, the actual history $E$ is not linearizable, but the detected history $E'$ is linearizable
due to a long delay between the event that announces the operation of \m{A} that $p_1$ is going to call next (Line~\ref{L05})
and the actual moment when the operation is called (Line~\ref{L06}).
}
\label{fig-stretch}
\end{center}
\end{figure}

The impossibility in Theorem~\ref{theo-impossibility} comes from the inability of the processes in any verifier \m{V} to
detect the actual history of an arbitrary implementation \m{A}. The history is ultimately defined 
by the order of occurrence of local events of processes (invocations and responses). 
Since the processes are asynchronous, 
a delay of arbitrary length can happen between the steps of a process in Lines~\ref{L05} and~\ref{L06}, 
and between the steps in Lines~\ref{L07} and~\ref{L08}, in the generic verifier in Figure~\ref{fig-verifier}.
Thus, basically the processes are only able to detect a history of~\m{A} where the operations might be ``stretched''.
Figure~\ref{fig-stretch} schematizes this situation when \m{A} is a queue.
Essentially, this is the main argument in the proof of Theorem~\ref{theo-impossibility}.

Let $E$ denote the actual history of $\m{A}$ and $E'$ denote the history of \m{A} detected by the processes.
Hence, the history~$E'$ only ``sketches'' the actual history $E$.
Actually, it can be that $E'$ is linearizable despite $E$ being non-linearizable,
as some operations that do not overlap in $E$ might actually overlap in~$E'$ 
(see the example at the bottom in Figure~\ref{fig-stretch}).
Thus, we only have the following implication:
$$E \hbox{ is linearizable } \, \Longrightarrow \, E' \hbox{ is linearizable}.$$

Arguably, this is the best the processes can do in any verifier \m{V}
for detecting the actual history of \m{A}. For the time being, let us suppose that 
the processes somehow can compute~$E'$, and are able to write it to and read it from the shared memory. 
With the help of~$E'$, the processes can easily satisfy soundness in a verifier,
each process simply needs to locally test whether $E'$ is linearizable.
Completeness however cannot be satisfied: 
if $E$ is not linearizable, $E'$ might or might not be linearizable 
(see again the example at the bottom in Figure~\ref{fig-stretch}).
This discussion suggests a weaker version of the distributed runtime verification problem, requiring 
soundness and a weaker version of completeness where processes are allowed to output \emph{false positives},
i.e.,~\error is not reported although the actual history of \m{A} is not linearizable.
This weaker version of the problem can be formally defined and shown to be solvable~\cite{thesis-valeria}.
However, this result is not completely satisfactory, since, after all, the main motivation
of any runtime verification technique is preventing incorrect responses.
We can do better, as explained next.

\begin{figure}[ht]
\begin{center}
\includegraphics[scale=0.65]{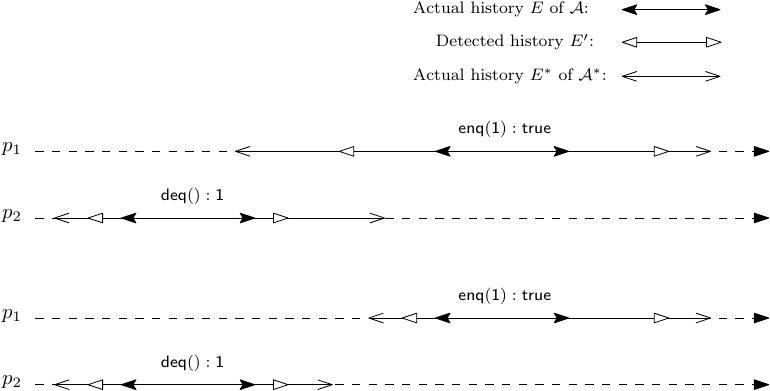}
\vspace{-0.1cm}
\caption{The operations of any execution $E^*$ of $\m{A}^*$ might ``shrink'' in the detected history $E'$. 
Two executions of $\m{A}^*$ where $\m{A}$ is a queue implementation
are shown; an operation $\Apply(\op)$ is simply denoted $\op$.
In the example at the top, the actual history of $\m{A}^*$ is linearizable but the history detected inside $\m{A}^*$ is not.
In contrast, in the example at the bottom, the actual history of $\m{A}^*$ is not linearizable, which implies that
the detected history is not linearizable either.
}
\label{fig-shrink}
\end{center}
\end{figure}

Above we only achieved soundness because we had:
$E \hbox{ is linearizable } \, \Longrightarrow \, E' \hbox{ is linearizable}$, namely, the implication goes from
the actual history to its sketch. If the implication is reversed, we can achieve completeness instead.
This is what we do, for an implementation obtained from~\m{A}.
We take the mechanism that computes $E'$ and use it to produce a \emph{new}
implementation $\m{A}^*$ that ``wraps''~\m{A}, where processes output 
the responses obtained from~\m{A}, somehow together with the detected history $E'$
(each process outputs only a ``piece'' of $E'$).
It turns out that $\m{A}$ and $\m{A}^*$ have the same progress properties, and
$\m{A}$ is linearizable if and only if $\m{A}^*$ is linearizable.
We now have that in every history $E^*$ of $\m{A}^*$,
$E'$ sketches $E^*$
with the difference that, compared to $E^*$, the operations in $E'$ might ``shrink''.
Two histories of $\m{A}^*$ when \m{A} is queue are schematized in Figure~\ref{fig-shrink}.
We thus have:
$$E' \hbox{ is linearizable} \, \Longrightarrow \, E^* \hbox{ is linearizable},$$ as desired.

Thus, if $E^*$ is not linearizable 
then its sketch $E'$ (which is part of $\m{A}^*$'s output) is not linearizable either,
and hence in a \emph{verifier for the derived implementation} $\m{A}^*$,
processes now can fulfill completeness  by locally analyzing $E'$ 
(in the verifier, processes need to exchange information to obtain the whole sketch~$E'$).
For soundness, however, it can be the case that  $E'$ is not linearizable although~$E^*$ is indeed linearizable 
(see the example at the top in Figure~\ref{fig-shrink}).
It turns out that $E'$ is a history of $\m{A}^*$ and then it \emph{witnesses} that $\m{A}^*$ is not linearizable.
Hence, processes have a reason to report \error: the current history of $\m{A}^*$
might be linearizable but the processes somehow have ``predicted'' that $\m{A}^*$ is not linearizable, 
and they have a history that witnesses that fact.
Moreover, no process can distinguish between $E^*$ and $E'$ (i.e. the histories are equivalent),
hence it is actually possible that the actual history of $\m{A}^*$ is $E'$ instead of~$E^*$.
Therefore, the processes in a verifier can fulfill a \emph{predictive} version of soundness for $\m{A}^*$, 
where they are allowed to report \error when $\m{A}^*$'s current history is linearizable,
as long as this comes with a witness.

The previous discussion motivates the predictive\footnote{{We have chosen to use `predictive' 
as the modified soundness property is reminiscent to predictive dynamic data-race detection
techniques (e.g.,~\cite{drp1, drp2, drp3, drp4}), which soundly report data-race violations by analyzing program executions,
even if the data-race violation is not observed in an execution.}} 
version of the runtime verification problem below,
which will be shown to be solvable for implementations such as $\m{A}^*$.
\RV will denote the class with the implementations like $\m{A}^*$.

\begin{definition}[Distributed Runtime Predictive Verification]
Let $\m{O}$ be a concurrent object specified in some way, and consider a correctness condition
${\sf P}_\m{O}$ for $\m{O}$. We say that a \emph{wait-free} verifier~$\m{V}_\m{O}$
\emph{distributed runtime predictively verifies}~${\sf P}_\m{O}$
if the following two requirements are fulfilled in every infinite execution $E$ of $\m{V}_\m{O}$
with an arbitrary input (abstract) implementation~$\m{A}^*$:
\begin{enumerate}
\item Predictive Soundness: If for every finite prefix $E'$ of $E$, $E'|\m{A}^*$ satisfies ${\sf P}_\m{O}$, 
then either no process reports \error, 
or at least one process reports \error together with a witness for $\m{A}^*$.

\item Completeness. If $E$ has a finite prefix $E'$ such that $E'|\m{A}^*$ does not satisfy~${\sf P}_\m{O}$, 
then at least one process reports \error together with a witness for $\m{A}^*$.
\end{enumerate}

We say that ${\sf P}_\m{O}$ is \emph{distributed runtime predictively verifiable} if there is a wait-free verifier 
that distributed runtime predictively verifies ${\sf P}_\m{O}$.
\end{definition}

A slight variant of the proof of Theorem~\ref{theo-impossibility} shows that 
the predictive version above is impossible for the general class of implementations
(see Appendix~\ref{sec-impossibility-predict}).

\paragraph{An intuition about the construction from $\m{A}$ to $\m{A}^*$}
The idea in $\m{A}^*$ is to make asynchrony an ``ally'' instead of an ``enemy''.
Suppose that in an execution $E^*$ of~$\m{A}^*$, the actual history $E$ of~\m{A} is not linearizable. 
If delays are short, then
$E'$ is not linearizable (see the example at the bottom of Figure~\ref{fig-shrink}), 
and hence the mistake can be detected.
If delays are long, then the sketch $E'$ is linearizable (hence $E^*$ too);
in this case the mistake cannot be detected, but $E'$ somehow ``enforces'' linearizability in~$E^*$.
Therefore, a crucial property of $\m{A}^*$ is that
the sketches it produces allow any client \m{C} that uses $\m{A}^*$ instead of \m{A},  
detect any possibly non-linearizable history of \m{A} that \emph{is not enforced corect} by $\m{A}^*$.
Thus, \m{C}'s computation can never be compromised by incorrect responses of~\m{A},
as they are either enforced in $\m{A}^*$ or detectable in \m{C}.

\section{The class of Distributed Runtime Verifiable (\RV) Implementations}
\label{sec-class-DRV}

This section introduces the class \RV and shows its main properties.
Roughly speaking, it will show that (1) every implementation can be easily transformed into its counterpart in \RV
and (2) every implementation in \RV provides a ``sketch'' of the current execution,
additionally to the outputs of the object it implements.
As we will see, the sketches are the key property that makes any of these implementations predictively verifiable.
Furthermore, the same result holds not only for linearizability, 
but for a class of objects, with its companion correctness condition, that generalize linearizability.
After defining the generalized class of objects, 
the section defines the class \RV and shows the main properties of the implementations in the class.

\subsection{Generalizing linearizability}

In the rest of the paper, similarly to~\cite{HW90},
an \emph{abstract object}, called just \emph{object} for simplicity, is defined as a set of well-formed finite histories.
The associated correctness condition is the \emph{membership predicate}.
Thus, a finite history of an implementation is \emph{correct} with respect to the object 
if the history belongs to the set specifying the object.
Then, an implementation is \emph{correct} with respect to the object if each of its finite histories 
belong the set specifying the object.

Linearizability can be alternatively stated through the abstract object formalism:.

\begin{remark}[Linearizability as abstract objects]
{For any sequential object \m{O}, consider the abstract object $\m{O}'$
with every finite history that is linearizable with respect to \m{O}.
Then, for any implementation \m{A}, it holds that \m{A} is linearizable with respect to \m{O}
if and only if \m{A} is correct with respect to $\m{O}'$ 
(i.e., every history of \m{A} belongs to $\m{O}'$).}
\end{remark}

In order to generalize linearizability, we define a partial order on the set
of operations, complete \emph{and} pending, of any history $E$ of an implementation; 
the relation is denoted $\prec_E$.
For any two operations $\op$ and $\op'$ in $E$,
we have that ${\op} \prec_E {\op}'$ if and only if $res({\op})$ precedes $inv({\op}')$ in $E$. 
Thus, the only difference with $<_E$ is that $\prec_E$ also relates pending operations.

\begin{definition}[Similarity between histories]
A finite history $E$ is \emph{similar} to a finite history~$F$
if there is a history~$E'$ such that:
\begin{enumerate}
\item $E'$ can be obtained from $E$ by appending responses to some pending operations
and removing invocations of some pending operations,
\item $E'$ and $F$ are equivalent, and
\item $\prec_{E'} \, \subseteq \, \prec_F$.
\end{enumerate}
\end{definition}
 
\begin{definition}[The class \GenLin]
The class \emph{\GenLin} contains every abstract object that is 
closed by prefixes and similarity.
Namely, if the set specifying the object contains $F$, then it also contains:
\begin{enumerate}
\item every prefix $E$ of $F$, and
\item every history $E$ that is similar to $F$.
\end{enumerate}
\end{definition}

We now show that linearizability {is contained} in the class \GenLin.
{Namely, for every sequential object \m{O}, 
the abstract object with all finite histories that are linearizable with respect to \m{O}, belongs to \GenLin.}

\begin{lemma}[Linearizability is {contained} in \GenLin]
\label{lemma-prefix-superset}
Let $F$ be any finite history that is linearizable with respect to some sequential object $\m{O}$ (i.e. a state machine).
Then,
\begin{enumerate}
\item every prefix $E$ of $F$ is linearizable with respect to $\m{O}$, and

\item every history $E$  that is similar to $F$ is linearizable with respect to~$\m{O}$.
\end{enumerate}
Therefore, the abstract object with all finite histories that are linearizable with respect to~$\m{O}$ is a member of \GenLin.
\end{lemma}

\begin{proof}
The prefix closure proof in~\cite{GR14}[Theorem 4] assumes a definition of linearizability that is slightly different 
than the one we use here; that proof however also holds in our case.
For completeness, we present the proof.
Let $F = E E'$. Consider any linearization $S$ of $F$.
Then, there is an extension $F'$ of $F$ such that $comp(F')$ and $S$ are equivalent
and $<_{comp(F')} \, \subseteq \, <_S$.
Let $S = S_E S_{E'}$, where $S_E$ is the shortest
prefix of $S$ with all complete operations in~$E$. 
We argue that $S_E$ is a linearization of $E$.
Let us observe first that $S_E$ does not have an operation whose invocation appears in~$E'$:
(1) by definition of $S_E$, the last operation $\op$ of $S_E$ is complete in $E$,
and (2) if $S_E$ has such an operation $\op'$, then $\op' <_{S_E} \op$,
which contradicts that $<_{comp(F')} \, \subseteq \, <_S$ because
we certainly have $\op <_{comp(F')} \op'$.
Now, let $E''$ be the extension of $E$ obtained by appending all response in $S_E$ to
pending operations in $E$.
It is not hard to see that $comp(E'')$ and $S_E$ are equivalent:
if there is a~$p_i$ such that $comp(E'')|{p_i} \neq S_E|{p_i}$, then simply $comp(F')$ and $S$ are not equivalent.
We now argue that $<_{comp(E'')} \, \subseteq \, <_{S_E}$, 
from which we conclude that indeed $S_E$ is a linearization of $E$.
Consider any $\op <_{comp(E'')} \op'$.
Since $comp(E'')$ and $S_E$ are equivalent, both $\op$ and $\op'$ appear in~$S_E$.
Moreover,~$S_E$~is a sequential history, and hence $<_{S_E}$ must relate $\op$ and $\op'$.
If both~$op$ and~$op'$ are complete in $E$,
then $\op <_E \op'$, hence $\op <_F \op'$, and consequently $\op <_{comp(F')} \op'$,
and as $<_{comp(F')} \, \subseteq \, <_S$, we have $\op <_S \op'$, which implies $\op <_{S_E} \op'$.
Consider now the case where at least one of $\op$ and $\op'$ are pending in $E$.
Note that if $\op$ is pending in $E$, then a response to it is appended in $E''$, 
and hence it cannot be the case that $\op <_{comp(E'')} \op'$.
Thus, $\op$ is complete in $E$, and $\op'$ is pending in $E$.
In $F'$, either a response to $\op'$ is appended, or no responses to it is appended
because~$\op'$ is complete in $F$. In any case, we have $\op <_{comp(F')} \op'$,
and hence $\op <_S \op'$, as $<_{comp(F')} \, \subseteq \, <_S$, 
which ultimately implies that $\op <_{S_E} \op'$.
Thus, we conclude that $<_{comp(E'')} \, \subseteq \, <_{S_E}$, 
and hence $S_E$ is a linearization of~$E$.

For the second claim, consider any history $E'$ obtained from $E$ 
by appending responses to some pending operations and removing invocations of some pending operations,
with $E'$ and $F$ being equivalent and $\prec_{E'} \, \subseteq \, \prec_F$.
Let $I$ and $R_E$ denote the sets with the invocations and responses removed and appended, respectively, to obtain $E'$. 
Consider any linearization $S$ of $F$.
Then, there is an extension $F'$ of $F$ such that $comp(F')$ and $S$ are equivalent
and $<_{comp(F')} \, \subseteq \, <_S$.
Let $R_F$ denote the set with the responses appended to obtain $F'$.
Let $E''$ be any extension of $E$ obtained by appending to it the responses in $R_E \cup R_F$.
We prove that $comp(E'')$ and $S$ are equivalent and $<_{comp(E'')} \, \subseteq \, <_S$,
from which follows that $S$ is a linearization of $E$.
As $E'$ and $F$ are equivalent, it is easy to see that $comp(E'')$ and $comp(F')$ are equivalent 
(just note that the invocations in $I$ do not appear in $comp(F')$),
and hence $comp(E'')$ and $S$ are equivalent.
Now, consider any $\op <_{comp(E'')} \op'$.
Observe that it cannot be that the response of $\op$ is in $R_E \cup R_F$.
Then, $\op$ is completed in $E$; however $\op'$ might be pending or complete in $E$.
Note that $\op <_{comp(E'')} \op'$ implies $\op \prec_{E'} \op'$, 
and hence $\op \prec_{F} \op'$, as $\prec_{E'} \, \subseteq \, \prec_F$,
from which follows $\op <_{comp(F')} \op'$. 
We thus have $<_{comp(E'')} \, \subseteq \, <_{comp(F')}$,
and then $<_{comp(E'')} \, \subseteq \, <_S$ because $<_{comp(F')} \, \subseteq \, <_S$.
Therefore, $S$ is a linearization of $E$.
\end{proof}

Similarly, it can be shown that variants of linearizability such as 
{set-linearizability}~\cite{N94} and {interval-linearizability}~\cite{CRR18},
among others, are contained in the class \GenLin. 
The reason is that these variants differ from linearizability only in the properties of the linearization $S$
of a given execution. In some cases, $S$ can ``deviate'' from sequential executions of
state machine $\m{O}$, 
or it might be the case  that $S$ is not necessarily a sequential execution, 
where several operations can occur simultaneously at the same time, 
or even an operation can overlap several operations.
To adapt the proof above for set-linearizability and interval-linearizability, 
we only need to replace the sequential execution $S$
with a set-sequential and interval-sequential execution, respectively.

\subsection{The class $\RV$}

Let $\m{A}$ be \emph{any} implementation. 
We already have seen that it is impossible to runtime verify
linearizability of $\m{A}$ with respect to some sequential objects. 
We will now see that $\m{A}$ can be \emph{indirectly}
verified through an implementation $\m{A}^*$ that can be easily obtained from~$\m{A}$.
The implementation~$\m{A}^*$ appears in Figure~\ref{figure-A-star},
in which processes communicate though a linearizable snapshot object.

\begin{figure}[ht]
\centering{ \fbox{
\begin{minipage}[t]{150mm}
\scriptsize
\renewcommand{\baselinestretch}{2.5} \resetline
\begin{tabbing}
aaaa\=aaa\=aaa\=aaa\=aaa\=aaa\=aaa\=\kill 

{\bf Shared Variable:}\\

$~~$  $N$ = wait-free linearizable snapshot object, with each entry initialized to $\emptyset$\\ \\

{\bf Local Persistent Variable:}\\

$~~$  $set_i$ = a set initialized to $\emptyset$\\ \\

{\bf Operation}  $\Apply(\op_i$) {\bf is} \\

\line{N01} \> $set_i \leftarrow set_i \cup \{ (p_i, \op_i) \}$\\

\line{N02} \> $N.\W(set_i)$\\

\line{N03} \> Invoke operation $\Apply(\op_i)$ of \m{A}\\

\line{N04} \> $y_i \leftarrow$ response from operation $\Apply(\op_i)$ of \m{A}\\

\line{N05} \> $s_i \leftarrow N.{\Snap}()$\\

\line{N06} \> $\lambda_i \leftarrow \bigcup_{k \in \{1,2, \hdots, n\}} s_i[k]$\\

\line{N07} \> {\bf return} $(y_i, \lambda_i)$\\

{\bf end} \Verify

\end{tabbing}
\end{minipage}
  }
\caption{\small From $\m{A}$ to $\m{A}^* \in \RV$ (code of process $p_i$).}
\label{figure-A-star}
}
\end{figure}

\begin{definition}[The snapshot object~\cite{AADGMS93}]
{\emph{Snapshot} is a sequential object whose state 
is a shared array $MEM$ with $n$ entries, one per process, that can be modified
through two operations:
$\Snap()$ that returns an $n$-array with the content of all entries of $MEM$,
and $\W(v)$ that writes~$v$ in $MEM[i]$, where $i$ is the index of the process that invokes 
the operation.\footnote{Recall that the index of $p_i$ is $i$.}
The initial state of the snapshot object~is~$[\emptyset, \hdots, \emptyset]$.}
\end{definition}

There are wait-free linearizable snapshot implementations that use only \R/\W base objects,
e.g.~\cite{AADGMS93, IC94}. Thus, the steps in Lines~\ref{N02} and~\ref{N05} can be assumed to be atomic,
due to the modular properties of linearizability~\cite{HW90, SHP21}.
In the step complexity analysis in this section and the next one, we consider the
implementation in~\cite{IC94}, whose step complexity is $O(n)$.
{There have been proposed snapshot implementations whose 
theoretical step complexity is worse but with good performance in real-world system (e.g.~\cite{WBBFRS21})}.

In $\m{A}^*$, every process $p_i$ simply announces in a shared memory the next high-level operation~$\op_i$ 
it wants to execute, then obtains a response for $\op_i$ using $\m{A}$, atomically reads all operations
that have been announced in the snapshot object so far, storing them all together in a (unordered) set~$\lambda_i$,
and finally returns the set together with the response obtained from $\m{A}$. 
The set $\lambda_i$ is called the \emph{view} of $(p_i, \op_i)$.
As we will see, this simple mechanism, the views, succinctly 
encode a ``sketch'' of $\m{A}^*$'s current history,
and this sketch is what makes $\m{A}^*$ predictively verifiable.

{The construction in Figure~\ref{figure-A-star} uses \m{A} as a black-box,
hence it is not based on any property of \m{A}. Therefore, for \emph{any} implementation \m{A}, regardless of the
object it implements, one can construct~$\m{A}^*$ as described in Figure~\ref{figure-A-star}.}

\begin{definition}[The class of Distributed Runtime Verifiable Implementations]
\RV~denotes the class of concurrent implementations obtained through the
construction in Figure~\ref{figure-A-star}.
\end{definition}

\subsection{Analyzing $\m{A}^*$}

\subsubsection{$\m{A}^*$ preserves \m{A}'s properties}

We first show that $\m{A}^*$ preserves progress and correctness properties of $\m{A}$.
In the rest of the section, we disregard the views in the responses of~$\m{A}^*$ (i.e. the sets $\lambda_i$), 
unless stated otherwise.
Recall that the execution itself of an implementation also denotes the history obtained from~it.

\begin{lemma}[Correctness of $\m{A}^*$]
\label{lemma-from-A-to-A-star}
Consider any implementation \m{A} and any object $\m{O}$ in the class~\GenLin.
Then, $\m{A}$ is correct with respect to $\m{O}$ if and only if 
$\m{A}^*$ is correct with respect to $\m{O}$ (disregarding the views in the responses of $\m{A}^*$). 
Furthermore, $\m{A}^*$ is lock-free (resp. wait-free) if and only if  $\m{A}$ is lock-free (resp. wait-free), 
and the step complexity of $\m{A}^*$ is the step complexity of $\m{A}$ plus~$O(n)$.
\end{lemma}

\begin{proof}
First, it is easy to see the segment of code in Lines~\ref{N01} to~\ref{N02} is wait-free,
as well as the segment in Lines~\ref{N05} to~\ref{N07}.
Thus, if $\m{A}$ is lock-free (resp. wait-free) then $\m{A}^*$ lock-free (resp. wait-free), and vice versa.
As for step complexity, it executes one \W before invoking $\m{A}$,
and one \Snap after, which can be implemented in $O(n)$ step using the algorithm in~\cite{IC94}.
For correctness, we prove each direction separately.
\begin{itemize}
\item[$\Rightarrow$] Let $E$ be any finite execution of $\m{A}^*$. 
We show that $E$ is correct, namely, the history obtained from it (denoted $E$ as well) belongs to~$\m{O}$.
Since $E|{\m{A}} \in \m{O}$ and \m{O} is closed by similarity, both by assumption,
it suffices to prove that~$E$ is similar to history $E|{\m{A}}$.
Consider the history $E'$ obtained from $E$ by:
(1) appending to $E$ the response in $E|{\m{A}}$ to every operations that is pending in $E$
but complete in $E|{\m{A}}$, and
(2) removing every invocation of a pending operation in $E$ that does not appear in $E|{\m{A}}$.
From the definition of $E'$ and the pseudocode in Figure~\ref{figure-A-star}, it can be easily verified that 
$E'$ and $E|{\m{A}}$ are equivalent.
Additionally, it holds that $\prec_{E'} \, \subseteq \, \prec_{E|{\m{A}}}$:
if ${\op} \prec_{E'} {\op}'$, then we must have that $res({\op})$ precedes $inv({\op}')$ in $E|{\m{A}}$
because, in $\m{A}^*$, operation calls to \m{A} are nested in operation calls of $\m{A}^*$,
and hence ${\op} \prec_{E|{\m{A}}} {\op}'$.

\item[$\Leftarrow$] 
Let $E$ be any finite history of $\m{A}$. 
We argue that $E \in \m{O}$.
Asynchrony in the model guarantees the existence
of the following execution $E'$ of $\m{A}^*$:
\begin{enumerate}
\item for every process $p_i$, for each of its operations $\Apply(\op_i)$,
the corresponding invocation and the steps from Lines~\ref{N01} to~\ref{N02} appear all together right before
the invocation to $\Apply(\op_i)$ of \m{A} in Line~\ref{N02}, and 
\item for every process $p_i$, for each of its operations $\Apply(\op_i)$,
the corresponding response and the steps from Lines~\ref{N05} to~\ref{N07} appear all together right after
the response from $\Apply(\op_i)$ of \m{A} in Line~\ref{N04} (if there is one), and

\item $E$ and $E'$ are the same history.
\end{enumerate}

Basically, ${E'}$ is obtained from $E$ by adding steps of $\m{A}^*$ right before and after the invocations 
and responses in $E$.
By assumption, $\m{A}^*$ is correct with respect to $\m{O}$, and hence ${E'} \in \m{O}$,
from which follows that ${E} \in \m{O}$, as $E$ and $E'$ are the same history.
\end{itemize}
\end{proof}

\begin{figure}[ht]
\begin{center}
\includegraphics[scale=0.65]{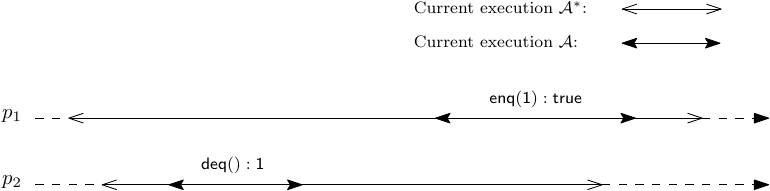}
\vspace{-0.1cm}
\caption{Due to asynchrony, $\m{A}^*$ is able to ``fix'' some histories of $\m{A}$. 
The figure depicts a history of  $\m{A}^*$ where the history of $\m{A}$ is not linearizable with respect to the queue,
however, the history of $\m{A}^*$ is linearizable.
Each operation $\Apply(\op)$ is simply denoted $\op$
and the views of the operations of $\m{A}^*$ are not shown.}
\label{fig-fixed-execution}
\end{center}
\end{figure}

As already explained at the end of Section~\ref{sec-overview},
$\m{A}^*$ can be alternatively understood as a mechanism 
that ``enforces'' correctness in \emph{some} incorrect histories of~$\m{A}$,
as the example in Figure~\ref{fig-fixed-execution} shows. 
Nevertheless, Lemma~\ref{lemma-from-A-to-A-star} implies that 
$\m{A}^*$ cannot enforce correctness in \emph{all} incorrect histories of~$\m{A}$.
For those histories that it is not able to ``fix'', the views provide 
a mechanism to detect they are incorrect,
which will be crucial to fulfill the completeness requirement of 
the distributed runtime predictive verification problem.

\subsubsection{Tight executions.}

\begin{definition}[Tight executions]
Let $\m{A}^*$ be any implementation in the class $\RV$, and consider any finite execution $E$ of it.
We say that $E$ is \emph{tight} if:
\begin{enumerate}
\item each of its pending operations has its \W step in Line~\ref{N02} but no \Snap step in Line~\ref{N05},
\item the invocation and local step in Line~\ref{N01} of every operation, complete or pending,
appear in a sequence right before its \W step in Line~\ref{N02}, and
\item the local steps in Lines~\ref{N06} and~\ref{N07} and response of every complete operation
appear in a sequence right after its \Snap step in Line~\ref{N05}.
\end{enumerate}
\end{definition}

Basically, in a tight execution, the beginning and end of an operation are identified
with the \W and \Snap steps in Lines~\ref{N02} and~\ref{N05}, respectively.

Any finite execution $E$ of $\m{A}^*$ can be transformed into a tight execution $T(E)$ of $\m{A}^*$ as described next:
\begin{enumerate}
\item remove the invocation and local step in Line~\ref{N01} of every pending operation with no \W step in Line~\ref{N02},
\item  for each of the remaining operations, complete or pending, move forward the invocation and local step in Line~\ref{N01}
 in a sequence right before its \W step in Line~\ref{N02},
\item for each of the remaining operations with its \Snap step in Line~\ref{N05}, 
first move backwards the local steps in Lines~\ref{N06} and~\ref{N07} 
and response (if there are any) in a sequence right after its \Snap step in Line~\ref{N05}, and
then insert (if necessary) the missing local steps in Lines~\ref{N06} and~\ref{N07} 
and response to complete the sequence that makes the operation complete.
\end{enumerate}
We say that $T(E)$ is the \emph{tight execution associated} to~$E$.

Observe that indeed $T(E)$ is an execution of $\m{A}^*$:
all invocations, responses and steps that are moved, forward of backward, or inserted
are \emph{local} to processes, and hence it is immaterial when they occur, or if they occur. 
Moreover, in $E$ and $T(E)$, the operations obtain the same responses from~\m{A} and compute the same views, 
as the order of invocations to and responses from~\m{A} are not modified to obtain $T(E)$, 
neither the order of \W and \Snap steps.
Intuitively, the difference between $E$ and~$T(E)$ is that operations in $T(E)$ span a possibly
``shorter'' interval of~time.

\begin{lemma}[Tight executions and actual executions]
\label{lemma-tight-executions}
Let $E$ be any finite execution of $\m{A}^*$.
Then, $T(E)$ is an execution of $\m{A}^*$, and for any object \m{O} in \GenLin, 

$$E|\m{A} \in \m{O} \, \Longrightarrow \, T(E) \in \m{O} \, \Longrightarrow \, E \in \m{O}.$$
\end{lemma}

\begin{proof}
As already argued, $T(E)$ is an execution of $\m{A}^*$. 
We will show that (history) $T(E)$ is similar to (history) $E|\m{A}$, and (history) $E$ is similar to (history) $T(E)$.
These two facts will prove the implications because \m{O} is closed by similarity,
as it belongs to \GenLin.

We argue first that $E$ is similar to $T(E)$. 
Let $E'$ be the history obtained from $E$ by:
(1) removing the invocation of every pending operation that is removed from $E$ to obtain $T(E)$, and
(2) appending the response in $T(E)$ of every pending operation in $E$ whose response is inserted to obtain $T(E)$.
Basically, $E'$ is obtained following steps (a) and (c) in the construction from $E$ to $T(E)$.
Note that $E$ and $T(E)$ are equivalent.
Consider now any ${\op} \prec_{E'} {\op'}$.
Hence $res(\op)$ precedes $inv(\op')$ in $E'$. 
Observe that it cannot be that $res(\op)$ is appended to $E$ to obtain $E'$, and $T(E)$,
and hence $\op$ is complete in $E$; $\op'$ is complete or pending.
We must have that ${\op} \prec_{T(E)} {\op'}$ becase to obtain $T(E)$,
responses of complete operations might only moved backward and 
invocations of complete or pending operations might only moved forward.
Therefore we have $\prec_{E} \, \subseteq \, \prec_{T(E)}$,
from which follows that $E$ is similar to~$T(E)$.

We show now that $T(E)$ is similar to $E|\m{A}$.
Consider the history $E'$ obtained from $T(E)$ by:
(1)~removing every invocation of a pending operation in $T(E)$ that does not appear in $E|{\m{A}}$
(any such operation executes its \W step in Line~\ref{N02} but does not executes the invocation in Line~\ref{N03}), and
(2)~appending the response in $E|{\m{A}}$ to every operations that is pending in $T(E)$
but complete in $E|{\m{A}}$ (any such operation operation executes its response in Line~\ref{N04},
but does not execute its \Snap step in Line~\ref{N05}).
Observe that $E'$ and $E|{\m{A}}$ are equivalent.
It is also true that $\prec_{E'} \, \subseteq \, \prec_{E|{\m{A}}}$:
if ${\op} \prec_{E'} {\op}'$, then we must have that $res({\op})$ precedes $inv({\op}')$ in $E|{\m{A}}$
because in $\m{A}^*$ operation calls to \m{A} 
are nested between the \W and \Snap steps in Lines~\ref{N02} and~\ref{N05},
from which follows that ${\op} \prec_{E|{\m{A}}} {\op}'$.
Therefore, $T(E)$ is similar to~$E|\m{A}$.
\end{proof}

Lemma~\ref{lemma-tight-executions} suggests a way to predictively verify $\m{A}^*$, through its tight executions;
the idea is actually simple.
Suppose that somehow processes are able to compute $T(E)$ of any execution~$E$ of~$\m{A}$.  
First, Lemma~\ref{lemma-tight-executions} shows that $T(E)$ is a history of $\m{A}^*$.
If $E$ is not correct, then the lemma also implies that $T(E)$ is not correct either,
and hence it is a witness for $\m{A}^*$, which can be reported to satisfy completeness;
and if $E$ is correct, then $T(E)$ might be correct or not, but in either case predictive soundness can be satisfied
because if $T(E)$ is not correct, it is a witness for~$\m{A}^*$ that can be reported.

In the rest of the section, 
we argue that the views encode the tight execution associated to the actual execution of~$\m{A}^*$,
which opens the possibility to implement the simple idea just described.

\subsubsection{From views to tight executions and vice versa.}
\label{sec-from-views-to-executions}
\begin{figure}[ht]
\begin{center}
\includegraphics[scale=0.65]{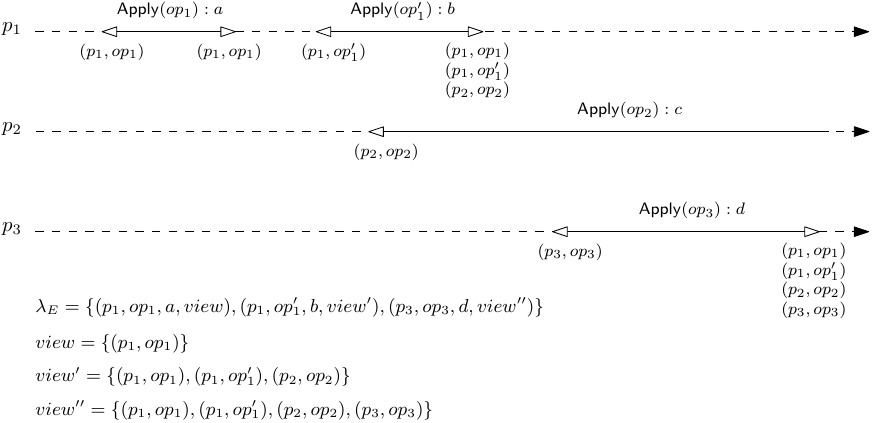}
\vspace{-0.1cm}
\caption{The figure shows the history of a tight execution $E$ of $\m{A}^*$. 
The invocation pair and the view of each operation is depicted.
The set $\lambda_E$ is depicted too. 
It is easy to check that the construction from $\lambda_E$ produces a history $X(\lambda_E)$ that is equivalent 
to $E$ with $\prec_E \, =  \, \prec_{X(\lambda_E)}$, and hence $E$ and $X(\lambda_E)$
are similar to each other.}
\label{fig-views-executions}
\end{center}
\end{figure}

Let $E$ be any finite execution of $\m{A}^*$.
In the associated tight execution~$T(E)$, 
let us replace each invocation 
of $p_i$ to operation $\Apply(\op_i)$, with the \emph{invocation pair} $(p_i, \op_i)$, and 
replace each response from operation $\Apply(\op_i)$ to $p_i$ (if there is one) with the set with all
invocation pairs $(p_j, \op_j)$ that precedes the response.
Figure~\ref{fig-views-executions} depicts an example of the replacement.
Since invocations and responses in $T(E)$ are associated to the \W and \Snap 
steps in Lines~\ref{N05} and~\ref{N02}, respectively, the view returned by any operation 
in $T(E)$ is exactly the set just defined. 
Below, we show that a ``sketch'' of the history $T(E)$ can be directly 
obtained from the views of operations in $E$.
First, from the specification of the linearizable \Snap object and 
the pseudocode of $\m{A}^*$, we obtain the following:

\begin{remark}[Properties of views]
\label{remark-snapshot-property}
Consider the views $\lambda_i$ and $\lambda_j$ in the responses
of any pair of completed operations $\Apply(\op_i)$ and $\Apply(\op_j)$ by $p_i$ and $p_j$ (possibly with $p_i = p_j$)
in any execution of $\m{A}^*$. The next properties are satisfied.
\begin{enumerate}
\label{remark-snapshot}
\item Self-inclusion: $(p_i, \op_i) \in \lambda_i$.
\item Containment comparability: $\lambda_i \subseteq \lambda_j \vee \lambda_j \subseteq \lambda_i$.
\item Process sequentiality: if $p_i = p_j \wedge \op_i \neq \op_j$, then 
$(p_i, \op_i) \notin \lambda_j \vee (p_j, \op_j) \notin \lambda_i$.
\end{enumerate}
\end{remark}

In the rest of this subsection, our discussion is focused on the tight executions of $\m{A}^*$,
particularly, in showing how the histories associated to them can be obtained from their views.

For any tight execution $E$ of $\m{A}^*$, 
let $\lambda_E$ denote the set with all 4-tuples $(p_i, \op_i, y_i, \lambda_i)$ 
such that $(y_i, \lambda_i)$ is the response of operation 
$\Apply(\op_i)$ of $p_i$ in $E$ (see Figure~\ref{fig-views-executions}).
We now explain that a well-formed history $X(\lambda_E)$
can be obtained from $\lambda_E$, and explain in what sense 
$X(\lambda_E)$ is a sketch of~$E$.
The construction that follows is from~\cite{CRR18}.

By Remark~\ref{remark-snapshot-property} (2), all \emph{distinct} views that appear in $\lambda_E$ can be ordered 
in strictly containment ascending order: $\sigma_1 \subset \sigma_2 \subset \hdots \subset \sigma_m$.
Let $\sigma_0$ denote $\emptyset$.
For each $k = 1, 2, \hdots, m$, $X(\lambda_E)$ is iteratively obtained following the
next two steps in ascending order (see Figure~\ref{fig-views-executions} for an example):
\begin{enumerate}
\item For each invocation pair $(p_i, \op_i) \in \sigma_k \setminus \sigma_{k-1}$,
the invocation to operation $\Apply(\op_i)$ by $p_i$ is appended to $X(\lambda_E)$;
the invocations are appended in any arbitrary order.

\item For each $(p_i, \op_i, y_i, \lambda_i) \in \lambda_E$ with $\lambda_i = \sigma_k$,
the response from operation $\Apply(\op_i)$ with output $(y_i, \lambda_i)$ by $p_i$ is appended to $X(\lambda_E)$;
the responses are appended in any arbitrary order. 
\end{enumerate}

In each of the steps of the construction above, 
either a set of invocations or responses are placed in some arbitrary sequential order.
For any of these orders, the resulting history has the 
same relation $\prec$
over pending and complete operations, by construction,
and hence all possible histories obtained in this way are similar to one another.
Thus, in fact, $X(\lambda_E)$ denotes an \emph{equivalence class} of histories.\footnote{In the parlance of~\cite{CRR18}, 
$X(\lambda_E)$ is an \emph{interval-sequential history}, i.e., 
an alternating sequence of non-empty sets with only either invocations or responses, starting
with a set of invocations. Interval-sequential histories are used in~\cite{CRR18} to define {interval-linearizability},
a generalization of linearizability in which, roughly speaking, an operation is allowed to overlap other
operations in an {interval-linearization} of an execution.} 
By similarity-closure of \GenLin, we have:

\begin{claim}
For every object $\m{O}$ in \GenLin and every tight execution $E$ of $\m{A}^*$,
either all histories of~$X(\lambda_E)$ are in $\m{O}$ or none of them is in $\m{O}$.
\end{claim}

By abuse of notation, we let $X(\lambda_E)$ denote any history of the equivalence class, unless stated otherwise.

The duality between histories and sets of views has been investigated in~\cite{CRR18}, 
where it is shown that the construction above 
is a bijection between (equivalence classes of) well-formed finite histories and sets of 4-tuples $(p_i, \op_i, y_i, \lambda_i)$
whose views satisfy the properties in
Remark~\ref{remark-snapshot}. 
The views can be understood as a \emph{static} mechanism,
{made of unordered sets}, that fully capture the dynamic real-time order
of operations in a history. Lemma 7.1 in~\cite{CRR18} directly implies that $X(\lambda_E)$ is 
indeed an accurate sketch of $E$ in the following sense: 

\begin{lemma}[Views are sketches of tight executions]
\label{lemma-views-executions}
For any tight execution $E$ of $\m{A}^*$,
$E$~and~$X(\lambda_E)$ are equivalent with $\prec_E \, = \, \prec_{X(\lambda_E)}$,
hence the histories are similar to one another.
\end{lemma}

As every $\m{O} \in \GenLin$ is closed by similarity, we have:

\begin{corollary}
\label{coro-views-exec-1}
For any tight execution $E$ of $\m{A}^*$ and every object $\m{O}$ in \GenLin, 
$$E \in \m{O} \Longleftrightarrow X(\lambda_E) \in \m{O}.$$
\end{corollary}

\begin{claim}
\label{claim-histo-implem}
If $H$ is a history of an implementation \m{B} (not necessarily in the class \RV), 
then any history $H'$ that is equivalent to $H$ 
with $\prec_H \, = \, \prec_{H'}$ is a history of \m{B} as well.
\end{claim}

\begin{proof}
Since $H$ is a history of \m{B}, there is an execution $E$ of \m{B} such that
$H$ is the history obtained from $E$.
History $H$ has the form $I_1 R_1 I_2 R_2 \hdots$,
where each $I_x$ (resp. $R_x$) is a non-empty sequence of invocations (resp. responses);
note that the invocations in $I_x$ (resp. responses in $R_x$) does not necessarily appear in
a continuos sequence in $E$.
The main observation to prove the claim is that for every history $H'$ 
such that is equivalent to $H$ with $\prec_H \, = \, \prec_{H'}$,
we must have that $H' = I'_1 R'_1 I'_2 R'_2 \hdots$ with 
$I'_x$ (resp $R'_x$) being a permutation of $I_x$ (resp. $R_x$).
Then, consider the execution $E'$ obtained from~$E$ as follows:
(1) for each $I_x$, first move all invocations in $I_x$ to the position of the \emph{first} invocation in~$I_x$,
and then permute them according to~$I'_x$, and similarly 
(2) for each $R_x$, first move all responses in $R_x$ to the position of the \emph{last} response in~$R_x$,
and then permute them according to~$R'_x$.
Observe that $E'$ is an execution of $\m{B}$ because only invocation and responses of $E$, which are local steps, 
are modified to obtain $E'$, and it is immaterial when these local steps actually occur, as long
as the specification of \m{B} is satisfied, as it happens in $E'$. 
By construction, $H'$ is the history obtained from~$E'$,
and therefore, $H'$ is a history of~$\m{B}$.
\end{proof}

Finally, Lemma~\ref{lemma-views-executions} and Claim~\ref{claim-histo-implem} imply:

\begin{corollary}
\label{coro-views-exec-2}
For any tight execution $E$ of $\m{A}^*$, $X(\lambda_E)$ is a history of $\m{A}^*$.
\end{corollary}

We stress that all claims in this subsubsection but Claim~\ref{claim-histo-implem} focus on the tight executions of $\m{A}^*$.
In fact, Corollary~\ref{coro-views-exec-1} is not true if $E$ is not tight,
as the example at the top of Figure~\ref{fig-shrink} shows
($E$ is linearizable but $X(\lambda_E)$, which corresponds to $E'$ in the example,
is not).

\subsection{A note on the class \RV and refined task solvability}
The views mechanism was introduced in~\cite{CRR18} (implicitly defined in~\cite{G14} too)
 in order to extend the task specification formalism~\cite{HKR13}
 to make it able to capture linearizable sequential long-lived objects. The resulting formalism
is called {multi-shot refined tasks}, where processes are required to produce 
outputs and \emph{implicitly} produce views.
It turns out that multi-shot refined tasks are strictly more expressive than linearizability,
and are equally expressive as interval-linearizability, also introduced in~\cite{CRR18}:
for every interval-sequential object, there is an equivalent multi-shot refined task, and vice versa.
From this perspective, the implementations in \RV solve the corresponding equivalent multi-shot refined tasks,
producing \emph{explicit} views, which is what makes them predictively runtime verifiable.

\section{Predictive runtime verifiability of \RV and self-enforced \GenLin implementations}
\label{sec-DRV-verifiable}

\subsection{A wait-free predictive verifier for \RV}
Lemmas~\ref{lemma-tight-executions} and~\ref{lemma-views-executions} in the previous section are 
the basis of the wait-free predictive verifier $\m{V}_\m{O}$ in Figure~\ref{fig-verifier-RV}. 
The idea of the verifier is simple.
For any finite execution~$E$ of the verifier, 
the views in $\lambda_{T(E|{\m{A}^*})}$ sketch the tight execution $T(E|{\m{A}^*})$ associated to 
the current execution $E|{\m{A}^*}$ of $\m{A}^*$~(Lemma~\ref{lemma-views-executions}),
and tight executions suffice to fulfill predictive soundness and completeness~(Lemma~\ref{lemma-tight-executions}).
Thus, in $\m{V}_\m{O}$ the processes simply exchange their views using \W and \Snap
(Lines~\ref{M07} to~\ref{M09}) and then each process locally tests if the execution it reads from 
the shared memory is correct (Lines~\ref{M10} to~\ref{M12}).

\begin{figure}[ht]
\centering{ \fbox{
\begin{minipage}[t]{150mm}
\scriptsize
\renewcommand{\baselinestretch}{2.5} \resetline
\begin{tabbing}
aaaa\=aaa\=aaa\=aaa\=aaa\=aaa\=aaa\=\kill 

{\bf Shared Variables:}\\

$~~$  $M$ = wait-free linearizable snapshot object, with each entry initialized to $\emptyset$\\ \\

{\bf Operation}  $\Verify(\m{A}^* \in \RV)$ {\bf is} \\

\line{M01} \> $res_i \leftarrow \emptyset$\\

\line{M02} \> {\bf while} {\sf true} {\bf do}\\

\line{M03} \>\> $op_i \leftarrow$ non-deterministically chosen operation that does not appear in $res_i$ \\

\line{M04} \>\> Invoke operation ${\Apply}(op_i)$ of $\m{A}^*$\\

\line{M05} \>\> $(y_i, \lambda_i) \leftarrow$ response from operation ${\Apply}(op_i)$ of $\m{A}^*$\\

\line{M06} \>\> $res_i \leftarrow res_i \cup \{ (p_i, op_i, y_i, \lambda_i) \}$\\

\line{M07} \>\> $M.\W(res_i)$\\

\line{M08} \> \> $s_i \leftarrow M.{\Snap}()$\\

\line{M09} \> \> $\tau_i \leftarrow \bigcup_{k \in \{1,2, \hdots, n\}} s_i[k]$\\

\line{M10} \>\> {\bf if} $X(\tau_i) \notin \m{O}$ {\bf then} \\

\line{M11} \>\> \> {\bf {\bf report}} $(\error, X(\tau_i))$\\

\line{M12} \>\> {\bf end if} \\

\line{M13} \> {\bf end while}\\

{\bf end} \Verify

\end{tabbing}
\end{minipage}
  }
\caption{\small A wait-free predictive verifier $\m{V}_\m{O}$ with linear step complexity, 
for any object $\m{O} \in \GenLin$
and any implementation $\m{A}^* \in \RV$
(code of process $p_i$).}
\label{fig-verifier-RV}
}
\end{figure}

Although $\m{V}_\m{O}$ relies on a simple idea, proving it correct is not simple at all.
The main reason is that in its \Snap step in Line~\ref{M08}, a process might obtain only a proper subset of 
 $\lambda_{T(E|{\m{A}^*})}$, hence the history $X(\tau_i)$, locally constructed as explained in Section~\ref{sec-from-views-to-executions},
 used in the test in Line~\ref{M10} might not be exactly~$T(E|{\m{A}^*})$, the sketch of the current history of $\m{A}^*$.
This situation might happen due to asynchrony, as some processes might have already obtained a response from
$\m{A}^*$ in $E|{\m{A}^*}$ (Line~\ref{M05}) but no written yet their responses in $M$ (Line~\ref{M07}).
Namely, processes might make decisions (Line~\ref{M10}) with incomplete information.
The main issue one needs to resolve when proving the correctness of $\m{V}_\m{O}$ is that
processes make consistent decisions, despite the possible lack of information.
To deal with this issue,
Lemmas~\ref{lemma-views-asyn} and~\ref{lemma-views-prefix} below show useful  
properties of the histories computed in Line~\ref{M10}.

As mentioned in the previous section, the \Snap object can be wait-free linearizable implemented using only \R/\W base objects,
and moreover, there are implementations with $O(n)$ step complexity~\cite{IC94}.
Thus, clearly $\m{V}_\m{O}$ is a wait-free, and uses only \R/\W base objects,
and every iteration of the while loop takes $O(n)$ steps.
Thus we have:

\begin{claim}
\label{claim-verifier-progress}
The verifier $\m{V}_\m{O}$ in Figure~\ref{fig-verifier-RV} is wait-free,
uses only \R/\W base objects with step complexity $O(n)$.
\end{claim}

For proving correctness of $\m{V}_\m{O}$, we analyze only its infinite executions
in which the sequence of \emph{local} computation steps in Lines~\ref{M09} to~\ref{M12} of a process $p_i$
appear all together right after the previous \Snap step of $p_i$ in Line~\ref{M08} (which is part of the same loop iteration).
Restricting our attention to these executions facilitates the discussion and proves the verifier to be correct in all cases,
as it is immaterial when these local steps actually occur.
This restriction can also be seen from a slightly different perspective: given any infinite execution of the verifier, 
the local steps in Lines~\ref{M09} to~\ref{M12} of a process $p_i$ can be moved backwards 
to be right next to the previous \Snap step of $p_i$ in Line~\ref{M08},
and the processes still make the same decisions in the modified execution.

\begin{lemma}
\label{lemma-views-asyn}
Let $E$ be any infinite execution of the wait-free verifier $\m{V}_\m{O}$ in Figure~\ref{fig-verifier-RV}
with an arbitrary input $\m{A}^* \in \RV$. Consider any  finite prefix $E'$ of $E$
whose last sequence of steps correspond to the steps in Lines~\ref{M08} to~\ref{M12} of a process $p_i$,
and let $\tau'$ denote the content of $\tau_i$ of $p_i$ at the end of~$E'$.
Then:
\begin{enumerate}
\item $X(\tau')$ is a history of $\m{A}^*$;
\item for every object $\m{O}$ in \GenLin, $T(E'|\m{A}^*) \in \m{O} \Longrightarrow X(\tau') \in \m{O}$.
\end{enumerate}
\end{lemma}

\begin{proof}
By Lemma~\ref{lemma-tight-executions}, $T(E'|\m{A}^*)$ is an execution of $\m{A}^*$.
By analyzing $T(E'|\m{A}^*)$, we will conclude that $X(\tau')$ is a history of $\m{A}^*$.
By the definition of $E'$, it follows that $\tau'$ contains all 4-tuples that appear in $M$ at the end of~$E'$.
Due to asynchrony, it is possible that not all 4-tuples in $\lambda_{T(E'|\m{A}^*)}$
appear in $M$ at the end of~$E'$; the reason is that it can be the case that in $E'$ a process 
executes its \Snap in Line~\ref{N05} of $\m{A}^*$, and hence the 4-tuple of the corresponding operation
is in $T(E'|\m{A}^*)$, by definition of tight executions, but it does not execute its \W step
in Line~\ref{M07} of $\m{V}_\m{O}$. 
Hence, at the end of $E'$, the response of the last operation in $T(E'|\m{A}^*)$
of a process in might be ``missing'' in~$\tau'$. 
We thus have that $\tau' \subseteq \lambda_{T(E'|\m{A}^*)}$;
moreover, $\lambda_{T(E'|\m{A}^*)} \setminus \tau'$ has at most one 4-tuple for each process.

We will modify $T(E'|\m{A}^*)$ to obtain an execution $F$ of $\m{A}^*$ whose history
is precisely $X(\tau')$.
As already said, each 4-tuple that appears in $\lambda_{T(E'|\m{A}^*)} \setminus \tau'$ corresponds to
an operation that is complete in $T(E'|\m{A}^*)$ and that operation is the last one of 
the corresponding process in the execution.
Let $F$ be the execution obtained from $T(E'|\m{A}^*)$ as follows:
for each 4-tuple in $\lambda_{T(E'|\m{A}^*)} \setminus \tau'$, 
remove the steps in Lines~\ref{N05} to~\ref{N07} of $\m{A}^*$ (see Figure~\ref{figure-A-star})
of the operation corresponding to the 4-tuple.
Clearly, $F$ is an execution of $\m{A}^*$ as only the last steps and response of the last operation 
of some processes are removed from $T(E'|\m{A}^*)$.
Moreover, it is a tight execution as each of its pending operations does not execute the \Snap step 
in Line~\ref{N05} of $\m{A}^*$.
By construction, we have that $\tau' = \lambda_F$,
from which follows that $X(\tau')$ and $X(\lambda_F)$ are equivalent
with $\prec_{X(\tau')} \, = \, \prec_{X(\lambda_F)}$.
Finally, $F$ and $X(\lambda_F)$ are equivalent
with $\prec_{F} \, = \, \prec_{X(\lambda_F)}$, by Lemma~\ref{lemma-views-executions},
and hence $X(\tau')$ is a history of~$\m{A}^*$, by Claim~\ref{claim-histo-implem}
and as $F$ is an execution of $\m{A}^*$.

From the previous discussion we can see that $F$ is similar to $T(E'|\m{A}^*)$.
Let $F'$ be the history obtained from $F$ by appending the responses
in the 4-tuples of $\lambda_{T(E'|\m{A}^*)} \setminus \tau'$. 
Observe that $F'$ and~$T(E'|\m{A}^*)$ are equivalent, 
and $\prec_{F'} \, = \, \prec_{T(E'|\m{A}^*)}$.
Since any object $\m{O} \in \GenLin$ is closed by similarity, we conclude that 
$T(E'|\m{A}^*) \in \m{O} \Longrightarrow F \in \m{O}$.
From the discussion above, we know that $F$ and $X(\tau')$
are similar to one another, and hence $F \in \m{O} \Longleftrightarrow X(\tau')$,
and therefore $T(E'|\m{A}^*)~\in~\m{O}~\Longrightarrow~X(\tau')~\in~\m{O}$.
\end{proof}

\begin{lemma}
\label{lemma-views-prefix}
Let $E$ be any infinite execution of the wait-free verifier $\m{V}_\m{O}$ in Figure~\ref{fig-verifier-RV}
with an arbitrary input $\m{A}^* \in \RV$. Consider any finite prefix $E'$ of $E$.
There is a finite prefix $F$ of $E$ such that for every finite prefix $E''$ of $E$
that has $F$ as one its prefixes and whose last sequence of steps 
correspond to the steps in Lines~\ref{M08} to~\ref{M12} of a process $p_i$,
it holds that $T(E'|\m{A}^*)$ is similar to a prefix of $X(\tau'')$,
where $\tau''$ denotes the content of $\tau_i$ of $p_i$ at the end of $E''$.
\end{lemma}

\begin{proof}
Lemma~\ref{lemma-views-executions} implies that we can concentrate on 
$X(\lambda_{T(E'|\m{A}^*)})$ to reason about $T(E'|\m{A}^*)$.
Observe that due to asynchrony, it is possible that not all 4-tuples in $\lambda_{T(E'|\m{A}^*)}$
appear in $M$ at the end of $E'$; the reason is that it is possible that in $E'$ a process 
executes its \Snap in Line~\ref{N05} of $\m{A}^*$ (see Figure~\ref{figure-A-star}), 
and hence the 4-tuple of the corresponding operation
is in $\lambda_{T(E'|\m{A}^*)}$, by definition of tight executions, but it does not execute its \W step
in Line~\ref{M07} of $\m{V}_\m{O}$. 
Thus, at the end of~$E'$, the response of the last operation in $\lambda_{T(E'|\m{A}^*)}$
of a process might be ``missing'' in $M$. 
Since $E$ is infinite and by assumption fair (see Section~\ref{sec-model}), 
there is a finite prefix $F$ of it in which all 
4-tuples in $\lambda_{T(E'|\m{A}^*)}$ appear in $M$ at the end of $F$;
note that at the of $F$, 
$M$ might contain 4-tuples of operations that do not appear in $\lambda_{T(E'|\m{A}^*)}$.
We claim that $F$ is the prefix of $E$ with the desired property.

Let $E''$ be any finite prefix of $E$
that has $F$ as one its prefixes and whose last sequence of steps 
correspond to the steps in Lines~\ref{M08} to~\ref{M12} of a process $p_i$,
and let $\tau''$ denotes the content of $\tau_i$ of $p_i$ at the end of $E''$.
Note that $E'$ is a prefix of $F$.
Using similar arguments as in the proof of Lemma~\ref{lemma-views-asyn}, 
it can be argued that $\tau'' \subseteq \lambda_{T(E''|\m{A}^*)}$.
Moreover, the election of $F$ and the definition of tight execution give that
$\lambda_{T(E'|\m{A}^*)} \subseteq \tau''$.
It directly follows from the definition of tight execution that if an operation is pending 
in $T(E'|\m{A}^*)$, then in $E'$ it does not execute its \Snap step in Line~\ref{N05} of~$\m{A}^*$ (see Figure~\ref{figure-A-star}).
We thus have that the view of any 4-tuple in $\tau'' \setminus \lambda_{T(E'|\m{A}^*)}$
contains the largest view among the views in the 4-tuples in $\lambda_{T(E'|\m{A}^*)}$.

We will now reason how $X(\lambda_{T(E'|\m{A}^*)})$ and $X(\tau'')$ are constructed 
from $\lambda_{T(E'|\m{A}^*)}$ and $\lambda_{\tau''}$, respectively.
Let $\sigma'_1 \subset \sigma'_2 \subset \hdots \subset \sigma'_{k'}$ and 
$\sigma''_1 \subset \sigma''_2 \subset \hdots \subset \sigma''_{k''}$ be respectively 
the distinct views in $\lambda_{T(E'|\m{A}^*)}$ and $\tau''$, ordered in
ascending order by containement. 
For the reasons above exposed, we have that 
(1) $k' \leq k''$, (2) $\sigma'_{\ell} = \sigma''_{\ell}$, for $1 \leq \ell \leq k'$,
and (3) $\sigma'_{k'} \subset \sigma''_{\ell}$, for $k' < \ell \leq k''$.
Therefore, the construction of $X(\lambda_{T(E'|\m{A}^*)})$ and $X(\tau'')$
from $\lambda_{T(E'|\m{A}^*)}$ and $\tau''$ use the same
first $k'$ views.
Let $S$ be the subset of $\tau''$ with all 4-tuples whose views are 
subset of $\sigma'_{k'}$. 
Note that $\lambda_{T(E'|\m{A}^*)} \subseteq S$;
intuitively, the difference between $S$ and $\lambda_{T(E'|\m{A}^*)}$ is that 
$\lambda_{T(E'|\m{A}^*)}$ might be missing the response of the last operation of some processes and the views
of these operation are contained by $\sigma'_{k'}$.
Hence, $S \setminus \lambda_{T(E'|\m{A}^*)}$ has at most one 4-tuple for each process.
Let $H$ be the shortest prefix of $X(\lambda_{T(E'|\m{A}^*)})$ containing all operations whose 
views are subset of $\sigma'_{k'}$.
From the definition of $S$ and the construction of $X(S)$, it directly follows that $H$ and $X(S)$ are similar to each other.
To conclude the proof, we argue that $X(\lambda_{T(E'|\m{A}^*)})$ is similar to $H$.
Let $X'$ be the history obtained by appending to $X(\lambda_{T(E'|\m{A}^*)})$ the responses in the 4-tuples 
of $S \setminus \lambda_{T(E'|\m{A}^*)}$. 
From the discussion above, we can conclude that $X'$ and $H$ are equivalent
and $\prec_{X'} \, = \, \prec_H$.
Therefore, $X(\lambda_{T(E'|\m{A}^*)})$ is similar to~$H$,
and hence $T(E'|\m{A}^*)$ is similar to~$H$ too.
\end{proof}

We are finally ready to prove that \GenLin is predictively verifiable, with respect to the class \RV of implementations.

\begin{theorem}[\GenLin is runtime predictive verifiable with respect to the class \RV]
\label{theo-DRV-verifiable}
Let $\m{O}$ be any object in the class \GenLin. 
The verifier $\m{V}_\m{O}$ in Figure~\ref{fig-verifier-RV} is a wait-free predictive verifier 
for the correctness of $\m{O}$ for the class \RV of implementations.
Furthermore, $\m{V}_\m{O}$ satisfies the following properties.
\begin{enumerate}
\item Efficiency. It uses only \R/\W base objects with step complexity $O(n)$.

\item Soundness for correct executions of $\m{A}$. 
In each infinite execution $E$ of it with input $\m{A}^*$,
if for every finite prefix~$E'$ of~$E$, it holds that $E'|\m{A} \in \m{O}$
(i.e. the history of $\m{A}$, the underlying implementation of $\m{A}^*$, is correct),
then no process reports \error in $E$.

\item Stability. For every infinite execution $E$ of it in which at least one process reports \error,
there is a finite prefix of it such that it is reported \error in every new loop iteration starting after the prefix.
\end{enumerate}
\end{theorem}

\begin{proof}
As it is shown in Claim~\ref{claim-verifier-progress},
the verifier $\m{V}_\m{O}$ is wait-free, 
uses only \R/\W base objects with step complexity $O(n)$,
and hence it satisfies the efficiency property.

We now argue that $\m{V}_\m{O}$ is a predictive verifier for the correctness of $\m{O}$,
namely, it satisfies predictive soundness and completeness;
we also argue that it satisfies soundness for correct executions of~\m{A} and stability.
Consider any infinite execution $E$ of~$\m{V}_\m{O}$ with an
arbitrary input implementation~$\m{A}^* \in \RV$.

\begin{itemize}
\item Predictive soundness. Suppose that there is a finite prefix $E'$ of $E$
whose last sequence of steps correspond to the steps in Lines~\ref{M08} to~\ref{M12} of a process $p_i$,
and $p_i$ reports \error at the end of~$E'$.
Clearly, $p_i$ reports \error because $X(\tau') \notin \m{O}$, where 
$\tau'$ denotes the content of $\tau_i$ of $p_i$ at the end of $E'$;
hence $p_i$ reports the history $X(\tau')$.
By Lemma~\ref{lemma-views-asyn} (1), $X(\tau')$ is a history of $\m{A}^*$,
and hence is a witness for $\m{A}^*$. Thus, predictive soundness is satisfied.

\item Soundness for correct executions of $\m{A}$.
Consider any finite prefix $E''$ of $E$, 
whose last sequence of steps correspond to the steps in Lines~\ref{M08} to~\ref{M12} of a process $p_i$.
By Lemmas~\ref{lemma-tight-executions} and~\ref{lemma-views-asyn}~(2), 
$E''|\m{A} \in \m{O} \, \Longrightarrow \, T(E''|\m{A}^*) \in \m{O}  \, \Longrightarrow \, X(\tau'') \in \m{O}$,
where $\tau''$ denotes the content of $\tau_i$ of $p_i$ at the end of $E''$.
Thus, $p_i$ does not report \error at the end of $E''$.

\item Completeness and stability. 
Consider any finite prefix $E'$ of $E$ such that $E'|\m{A}^* \notin \m{O}$.
By Lemma~\ref{lemma-tight-executions},
$E'|\m{A}^* \notin \m{O} \, \Longrightarrow \, T(E'|\m{A}^*) \notin \m{O}$.
Lemma~\ref{lemma-views-prefix} guarantees the existence of 
a finite prefix $F$ of $E$ such that for every finite prefix $E''$ of $E$
that has $F$ as one its prefixes and whose last sequence of steps 
correspond to the steps in Lines~\ref{M08} to~\ref{M12} of a process $p_i$,
it holds that $T(E'|\m{A}^*)$ is similar to a prefix $H$ of $X(\tau'')$,
where $\tau''$ denotes the content of $\tau_i$ of $p_i$ at the end of $E''$.
Since \m{O} is closed by similarity,
we have $T(E'|\m{A}^*) \notin \m{O} \, \Longrightarrow \, H \notin \m{O}$,
and since it is closed by prefixes, 
$H \notin \m{O} \, \Longrightarrow \, X(\tau'') \notin \m{O}$.
Thus we conclude that $p_i$ reports $(\error, X(\tau''))$ at the end of $E''$.
By Lemma~\ref{lemma-views-asyn} (1), $X(\tau'')$ is a history of $\m{A}^*$,
and hence is a witness for~$\m{A}^*$.
Therefore, completeness is satisfied.
Observe that stability is satisfied too as the analysis holds for every 
such $E''$ having $F$ as a prefix.
\end{itemize}
\end{proof}

\subsection{Runtime self-enforced correct implementations for \GenLin}

From Theorem~\ref{theo-DRV-verifiable}, we can obtain a simple and generic methodology that,
starting with an arbitrary \GenLin implementation \m{A}, produces a \GenLin implementation
\m{B} whose responses are runtime verified. We call the implementations obtained
with this methodology \emph{self-enforced \GenLin}.
The methodology is as follows.

\begin{figure}[ht]
\centering{ \fbox{
\begin{minipage}[t]{150mm}
\scriptsize
\renewcommand{\baselinestretch}{2.5} \resetline
\begin{tabbing}
aaaa\=aaa\=aaa\=aaa\=aaa\=aaa\=aaa\=\kill 

{\bf Shared Variables:}\\

$~~$  $M$ = wait-free linearizable snapshot object, with each entry initialized to $\emptyset$\\ \\

{\bf Local Persistent Variable:}\\

$~~$  $set_i$ = a set initialized to $\emptyset$\\ \\

{\bf Operation}  $\Apply(\op_i$) {\bf is} \\

\line{H01} \> Invoke operation ${\Apply}(op_i)$ of $\m{A}^*$\\

\line{H02} \> $(y_i, \lambda_i) \leftarrow$ response from operation ${\Apply}(op_i)$ of $\m{A}^*$\\

\line{H03} \> $res_i \leftarrow res_i \cup \{ (p_i, op_i, y_i, \lambda_i) \}$\\

\line{H04} \> $M.\W(res_i)$\\

\line{H05} \> $s_i \leftarrow M.{\Snap}()$\\

\line{H06} \>  $\tau_i \leftarrow \bigcup_{k \in \{1,2, \hdots, n\}} s_i[k]$\\

\line{H07} \> {\bf if} $X(\tau_i) \in \m{O}$ {\bf then} \\

\line{H08} \>\> {\bf return} $y_i$\\

\line{H09} \> {\bf else}\\

\line{H10} \>\> {\bf return} $(\error, X(\tau_i))$\\

\line{H11} \> {\bf end if} \\

{\bf end} \Apply

\end{tabbing}
\end{minipage}
  }
\caption{\small {Self-enforced \GenLin implementation $\m{V}_{\m{O}, \m{A}}$ (code of process $p_i$).}}
\label{fig-self-enforced}
}
\end{figure}

Let \m{A} be any implementation for an object $\m{O} \in \GenLin$.
Consider the \RV implementation $\m{A}^*$ obtained from \m{A} (Figure~\ref{figure-A-star}), and
the predictive verifier $\m{V}_\m{O}$ for \m{O} (Figure~\ref{fig-verifier-RV}).
{Let $\m{V}_{\m{O}, \m{A}}$ be the implementation obtained 
from $\m{V}_\m{O}$ and $\m{A}^*$ as shown in Figure~\ref{fig-self-enforced}:
its high-level operation $\Apply(\op_i)$
executes Lines~\ref{M04} to~\ref{M09} of  $\m{V}_\m{O}$ with $\m{A}^*$,
and returns $resp_i$ if $X(\tau_i) \in \m{O}$,
and $(\error, X(\tau_i))$ otherwise.
We stress that the methodology is generic, 
as $\m{V}_{\m{O}, \m{A}}$ does not rely on any specific property~of~$\m{A}$.
}

Theorem~\ref{theo-runtime-verified} below shows that indeed $\m{V}_{\m{O}, \m{A}}$ self-enforces \GenLin correctness.
Furthermore, at any time $\m{V}_{\m{O}, \m{A}}$ is able to produce a history that 
\emph{certifies} that its responses so far are correct or incorrect.

\begin{theorem}[Runtime self-enforced correct implementations for \GenLin]
\label{theo-runtime-verified}
Let $\m{O}$ be any object in \GenLin and $\m{A}$ be \emph{any} implementation.
Consider the implementation $\m{V}_{\m{O}, \m{A}}$ obtained from~$\m{A}$ as specified in Figure~\ref{fig-self-enforced}.
Then,
\begin{enumerate}
\item  $\m{V}_{\m{O}, \m{A}}$ and \m{A} have the same progress condition,

\item if $\m{A}$ is correct with respect to $\m{O}$, then $\m{V}_{\m{O}, \m{A}}$ is correct with respect to $\m{O}$;
otherwise every finite execution of $\m{V}_{\m{O}, \m{A}}$  is correct with respect to $\m{O}$
up to a prefix (which does not necessarily exist) where every new operation returns \error
together with a witness for $\m{A}^*$,

\item if requested, $\m{V}_{\m{O}, \m{A}}$ can output a history that is similar to the history
of $\m{V}_{\m{O}, \m{A}}$ at the moment of the request.
\end{enumerate}
\end{theorem}

\begin{proof}
First, since $\m{V}_{\m{O}}$ is wait-free, by Theorem~\ref{theo-DRV-verifiable}, and
\m{A} and $\m{A}^*$ have the same progress condition, by Lemma~\ref{lemma-from-A-to-A-star},
we have that $\m{V}_{\m{O}, \m{A}}$ and \m{A} have the same progress condition.

Observe that, by the definition of $\m{V}_{\m{O}, \m{A}}$, 
every infinite execution $E$ of $\m{V}_{O}(\m{A}^*)$ is naturally mapped 
to a \emph{unique} infinite execution $E'$ of $\m{V}_{\m{O}, \m{A}}$,
where the histories of \m{A} and $\m{A}^*$ are exactly the same in both executions,
and every process passes through essentially the same sequence of local states.
Basically, $E'$ is obtained from $E$ by respectively replacing 
the beginning and end of a loop iteration with the invocation and response
of the operation invoked in the iteration.

Consider the case where $\m{A}$ is correct with respect to $\m{O}$,
and let $E$ be any infinite execution of~$\m{V}_{\m{O}, \m{A}}$.
The soundness for correct executions of $\m{A}$ property of  
$\m{V}_{\m{O}}$ (see Theorem~\ref{theo-DRV-verifiable}) implies that
no process reports \error in~$E$. Thus, every operation returns the same response in $E|\m{A}$ and~$E$.
Consider any finite prefix $E'$ of $E$.
Using a similar reasoning as in previous proofs, it can be shown that
$E'$ is similar to $E'|\m{A}$, and hence $E'|\m{A} \in \m{O} \, \Longrightarrow \, E' \in \m{O}$,
as \m{O} is closed by similarity.
Therefore, $\m{V}_{\m{O}, \m{A}}$ is correct with respect to $\m{O}$, 
as $\m{A}$ is correct with respect to $\m{O}$.

Suppose now that $\m{A}$ is not correct with respect to $\m{O}$,
and let $E$ be any infinite execution of~$\m{V}_{\m{O}, \m{A}}$.
Consider any finite prefix $E'$ of $E$.
The argument in the previous paragraph shows that if $E'|\m{A} \in \m{O}$,
then no process reports \error in $E'$, and hence $E' \in \m{O}$. 
Thus consider the case $E'|\m{A} \notin \m{O}$.
The completeness and stability properties of $\m{V}_{\m{O}}$ (see Theorem~\ref{theo-DRV-verifiable}) implies that eventually
all new operations in $E$ report \error together with a witness $X$ for $\m{A}^*$.
To conclude the proof, for the sake of contradiction, suppose that no process reports \error in $E'$ but $E' \notin \m{O}$.
Using a similar reasoning as in previous proofs, it can be shown that
$E'$ is similar to $T(E'|\m{A}^*)$, and hence $E' \notin \m{O} \, \Longrightarrow \, T(E'|\m{A}^*) \notin \m{O}$,
as \m{O} is closed by similarity.
Since no process reports \error in $E'$, all invocation an responses that appear in it
have been written in the shared memory $M$ at the end of $E'$.
Thus the last process that takes its \Snap in Line~\ref{M08} of $\m{V}_{\m{O}}$
reads the views of all these operations, and hence the history it computes
in Line~\ref{M08} is $T(E'|\m{A}^*)$, from which follows that it reports \error in $E'$,
as we already saw that $T(E'|\m{A}^*) \notin \m{O}$.
We have reached a contradiction.

Finally, at any time a process can snapshot from the shared memory of $\m{V}_{\m{O}}$
a history that is similar to $T(E'|\m{A}^*)$, which can be returned if requested through $\Apply$.
\end{proof}

At first glance, one might think that the self-enforced \GenLin
implementation $\m{V}_{\m{O}, \m{A}}$ in Theorem~\ref{theo-runtime-verified}
is somehow runtime verifying \m{A}, hence contradicting the impossibility in Theorem~\ref{theo-impossibility};
there is no contradiction however. As $\m{A}^*$ is able to enforce correctness in only \emph{some} incorrect executions of $\m{A}$
(see Sections~\ref{sec-overview} and~\ref{sec-class-DRV}), it is possible that there are executions of $\m{V}_{\m{O}, \m{A}}$ whose 
execution of $\m{A}$ is not correct, but no process reports \error, which happens because $\m{A}^*$
is able to fix that execution of $\m{A}$, i.e., it enforces a correct behavior.

{We conclude this subsection by observing that constructing $X(\tau_i)$ in Line~\ref{H07} of $\m{V}_{\m{O}, \m{A}}$
can be locally computed in polynomial time in the number of operations in $\tau_i$ 
(see the construction in Section~\ref{sec-from-views-to-executions}).
Thus, the local test in that line is computed in the time it takes to test if $X(\tau_i)$ is in~$\m{O}$, 
plus the polynomial overhead incurred to build $X(\tau_i)$.
This polynomial overhead is desirable for linearizability since it is known that, for some sequential objects,
linearizability of an execution can be decided in polynomial time~\cite{BEEH15, EE18}.}

\subsection{Concurrent systems with accountable and forensic guarantees}

{Consider a situation where a concurrent algorithm \m{C} uses an implementation \m{A} of an object $\m{O} \in \GenLin$.
Theorem~\ref{theo-runtime-verified} 
show that, in~$\m{C}$, \m{A} can be replaced with its self-enforced \GenLin version~$\m{V}_{\m{O}, \m{A}}$,
whose non-$\error$ responses are runtime verified. 
Moreover, if \m{A} is actually correct, then \m{C} never obtains $\error$ from $\m{V}_{\m{O}, \m{A}}$, hence
$\m{C}$ cannot distinguish it is using $\m{V}_{\m{O}, \m{A}}$ instead~of~\m{A}.
This idea can be extended to all \GenLin implementations used in \m{C}.
If all those implementations are replaced with their self-enforced \GenLin versions, the modified
algorithm \m{C} is enriched with a mechanism from which incorrect \GenLin implementation can be accounted,
and witnesses histories of the self-enforced \GenLin implementations can potentially be used in a forensic stage,
which~$\m{C}$ could trigger once an incorrect response is detected at runtime.}

\section{Extensions}
\label{sec-extensions}

\subsection{Base objects of bounded size}
The implementation $\m{A}^*$ (Figure~\ref{figure-A-star}) and the verifier $\m{V}_\m{O}$ (Figure~\ref{fig-verifier-RV})
use shared object of unbounded size.
This unrealistic assumption can be removed by representing sets as linked lists. 
For $\m{A}^*$, each entry in $N$ contains the first node of a single linked list with the
items in the set of the process associated to that entry. Every time a process adds an item to its
set, it creates a new node, links it to the first node of the list (which then becomes the second node)
and then writes the new node in its entry in $N$.
A snapshot operation in $\m{A}^*$ now returns a vector
of positions in the linked lists, hence the elements in the sets in the snapshot 
are the nodes that are accesible from the positions.
The sets in $\m{V}_\m{O}$ can be represented in the same~way. 

\subsection{Decoupled self-enforced \GenLin implementations}
\label{sec-decoupled}

{In the self-enforced implementation $\m{V}_{\m{O}, \m{A}}$ (Figure~\ref{fig-self-enforced}),
every process is in charge of producing responses and verifying them correct,
which, from a performance perspective, might be undesirable. 
This issue can be mitigated by \emph{decoupling} response production and verification.
Namely, a group of processes $v_1, \hdots, v_m$, the  \emph{verifiers}, is in charge of the verification task,
and the rest of the processes $p_1, \hdots, p_n$, the \emph{producers}, are in charge of obtaining
responses using $\m{A}^*$, and store the sketch produced by $\m{A}^*$ for the verifiers to
perform the verification task.
The decoupled version~$\m{D}_{\m{O}, \m{A}}$ of~$\m{V}_{\m{O}, \m{A}}$ appears in Figure~\ref{fig-decoupled}.
Observe that, in contrast to $\m{V}_{\m{O}, \m{A}}$,  
$\m{D}_{\m{O}, \m{A}}$ might produce incorrect responses 
(due to delays, at a given time verifiers might verify a response that was already returned), however,
eventually the verifiers detect them, assuming that not all of them crash in an execution.
In the terminology of~\cite{BFRT16}, the producers
implement the \emph{communication interface} and the verifiers implement the \emph{monitoring system}.
}

\begin{figure}[ht]
\centering{ \fbox{
\begin{minipage}[t]{150mm}
\scriptsize
\renewcommand{\baselinestretch}{2.5} \resetline
\begin{tabbing}
aaaa\=aaa\=aaa\=aaa\=aaa\=aaa\=aaa\=\kill 

{\bf Shared Variables:}\\

$~~$  $M$ = wait-free linearizable snapshot object, with each entry initialized to $\emptyset$\\ \\

{\bf Local Persistent Variable of Producer $p_i$:}\\

$~~$  $set_i$ = a set initialized to $\emptyset$\\ \\

{\bf Operation}  $\Apply(\op_i$) {\bf is} \, \, \%\% Operation for producer $p_i$ \\

\line{H01} \> Invoke operation ${\Apply}(op_i)$ of $\m{A}^*$\\

\line{H02} \> $(y_i, \lambda_i) \leftarrow$ response from operation ${\Apply}(op_i)$ of $\m{A}^*$\\

\line{H03} \> $res_i \leftarrow res_i \cup \{ (p_i, op_i, y_i, \lambda_i) \}$\\

\line{H04} \> $M.\W(res_i)$\\

\line{H05} \> {\bf return} $y_i$\\

{\bf end} \Apply \\ \\

{\bf Operation}  $\Verify()$ {\bf is} \, \, \%\% Operation for verifier $v_j$; starts at beginning of computation \\

\line{H06} \> {\bf while} {\sf true} {\bf do}\\

\line{H07} \> \> $s_j \leftarrow M.{\Snap}()$\\

\line{H08} \> \> $\tau_j \leftarrow \bigcup_{k \in \{1,2, \hdots, n\}} s_j[k]$\\

\line{H09} \>\> {\bf if} $X(\tau_j) \notin \m{O}$ {\bf then} \\

\line{H10} \>\>\> {\bf report} $(\error, X(\tau_j))$\\

\line{H11} \>\> {\bf end if} \\

\line{M12} \> {\bf end while}\\

{\bf end} $\Verify$ 

\end{tabbing}
\end{minipage}
  }
\caption{\small {Decoupled self-enforced \GenLin implementation $\m{D}_{\m{O}, \m{A}}$ (codes of producer $p_i$ and verifier $v_j$}).}
\label{fig-decoupled}
}
\end{figure}

\subsection{Verifying obstruction-free and blocking implementations, and task solvability}

The proposed interactive model can be modified to include obstruction-free~\cite{HS08} or blocking~\cite{HS08} implementations.
The difference is that the interaction between the implementation and a verifier 
might be finite because the implementation
might block; thus the required properties, predictive soundness and completeness, need to be satisfied
even if the interaction is finite. 

Similarly, task solvability can be predictively runtime verified.
A task is a one-shot distributed problem~\cite{HKR13}. It is known that any task can be equivalently modelled
and a one-shot interval-sequential interval-linearizable object~\cite{CRR18}, which belongs to \GenLin,
since \GenLin includes interval-linearizability.
Again, the difference is that the interaction with this object is finite. 
Moreover, since every process invokes exactly one high-level operation, 
now a process can declare that the computation is \emph{correct} if it reads from the shared memory
the views of all processes in the system and the corresponding history is correct.

\subsection{Extension to other models of computation}

The proposed implementations assume a fixed number of processes 
that can participate in the computation. 
The implementations can however be 
adapted to shared memory models with an unbounded number of processes, i.e., where there is no
prior knowledge on the number of processes that participate.
We only have to use the wait-free snapshot implementation in~\cite{GMT01} for this kind of models.

Due to the shared memory simulation algorithm in~\cite{ABD95},
all our algorithms, $\m{A}^*$, $\m{V}_\m{O}$, $\m{V}_{\m{O}, \m{A}}$ and $\m{D}_{\m{O}, \m{A}}$
can be simulated in 
asynchronous message-passing systems where less than half processes can crash in any execution.

\section{Related Work}
\label{sec-related-work}

\paragraph{Runtime verification}
Runtime verification is an active field of research with important advancements in the last two decades.
It has been employed in academia and industry for {verification}
before system deployment in order to ensure {correctness} after deployment.
Runtime verification has been used mostly to analyze software, however it has also been applied to
other types of systems, e.g. hardware, hybrid and embedded systems, cyber-physical systems, 
distributed and concurrent systems, financial transaction systems,
and more. For a detailed exposition of the field, we refer the reader to textbook~\cite{BF18}
and surveys~\cite{FHR13, HG05, LS09}.

\paragraph{Distributed runtime verification}
Distributed runtime verification of distributed systems
is considered and emergent and important topic, that poses several
challenges that are yet to be solved (see~\cite{BFRT16, E-HF18, F21, FPS18, LFKV18, SSABBCFFK19}).
Designing distributed, asynchronous, fault-tolerant communication interfaces 
are regarded as a difficult problem. 
As far as we know, there is no runtime verification algorithm
in the literature (for any correctness property) 
that is at the same time fully asynchronous and fault-tolerant. 
There have been proposed distributed runtime verification algorithms for a number of properties that are
failure-free synchronous message-passing
(e.g.~\cite{AFIMP20, SHKKNPPW12}),
fault-tolerant message-passing where processes have access to 
clocks that are synchronized at some level
(e.g.~\cite{BKZ15, BGKS20, BF16, E-HF20, FCF14, GXJLSBH22, KB18, RF22}),
and fully asynchronous message-passing or shared memory 
with failure-free processes (e.g.~\cite{CGNM13, FF08, FFY08, LR13, SS14, SVAR04}).

\paragraph{Distributed fault-tolerant runtime verification}
Our approach is close to
the series of papers~\cite{BFRR22, FRT13, FRT20} 
initiated by Fraigniaud, Rajsbaum and Travers, who pioneered the study of 
distributed fault-tolerant verification~\cite{FRT11}. 
In those works, it is studied a shared memory concurrent system that solves
a series of tasks~\cite{HKR13},
and the aim is to runtime verify that the outputs for each task are correct, 
i.e. they satisfy the inputs/outputs relation specifying the task.
To do so, an asynchronous wait-free read/write shared memory algorithm 
runs every time a task in the series is solved, and it is assumed that the verification algorithm
terminates before the processes solve the next task.
Thus, the distributed runtime verification algorithm proposed in those papers is distributed
and fault-tolerant but it is \emph{not} fully asynchronous. 
In sharp contrast, we consider a \emph{full} asynchrony and wait-free shared memory system,
where some processes might be verifying the current execution while at the same time 
others are executing a high-level operation.
\GenLin includes tasks, hence task solvability is covered by our results. 
The main concern in~\cite{BFRR22, FRT13, FRT20} is to understand how many 
distinct report values can be reported by the processes
(\emph{opinions} in the parlance of those papers)
in order to runtime verify an algorithm.
Our approach here is slightly different, as a process reports nothing
(or implicitly makes a ``so far so good'' report) as long as the computation looks correct
from its perspective, and, when a process ``sees'' enough information,
it can decide whether the computation is correct or not. 
Another crucial difference is that 
the wait-free interactive verifier proposed here can detect
real-time order violations (called validity for some problems) 
in a finer way, as real-time relations of the actual execution are taken into account,
whereas in~\cite{BFRR22, FRT13, FRT20} real-time order of events is not considered. 
For example, by observing only (input,output) pairs, for consensus it is impossible
to detect when a process ran solo and decided a value distinct from its input,
which violates validity. That scenario, in contrast, can be detected by our verifier
through the views mechanism of the class \RV.

\paragraph{Runtime verification of concurrent algorithms}
For concurrent shared memory algorithms, runtime verification has been mostly used 
to detect data races, serializability violations (also called atomicity violations) and deadlocks;
less studied properties are order instruction violations, 
missed signals, starvation, and high-level correctness conditions such as 
sequential consistency and linearizability (see~\cite{E-HF18, LFKV18, QT12}).
Typically, in these works, asynchronous failure-free processes are assumed,
and achieving a distributed communication interface is not a primary target.
Several algorithms have been proposed for detecting data races and serializability violations;
some algorithms use techniques based on 
the assumption that the concurrent algorithm under inspection uses locks
(e.g.~\cite{SBNSA97}), and others use some form of vector clocks to capture 
the happens-before relation, i.e. relations of causally-related events (e.g.~\cite{FFY08}), 
or a mixture of both (e.g.~\cite{FF08}).
None of these techniques can be adapted to our setting because:
(1) we focus on lock-free implementations and
(2) linearizability totally depends on the real-time order of non-causally related events.
In general, a main difficulty is capturing the actual execution of a concurrent algorithm,
as explained in~\cite[Section 4]{E-HF18}. 
A simple solution is to serialize events using a lock (e.g.~\cite{MJGCR12});
we find this type of solutions undesirable because, first, it might change the
progress condition of the algorithm under inspection (as explained in the Introduction), 
and second, the lock creates a bottleneck that compromises performance.
Other algorithms rely on bytecode-level added
instructions in order to capture the execution (e.g.~\cite{HR04}), and moreover there are 
dynamic analysis frameworks working at a bycodelevel that provide information of the current execution
for performing dynamic analysis (e.g.~\cite{FF10}). A problem with these techniques is that 
the moment when an event happens and the moment when the event is registered
are not the same (i.e. they do not occur atomically, as in our interactive model), 
and hence the actual execution might not be captured
(which is the main argument in the proof of the impossibility in Theorem~\ref{theo-impossibility}).

\paragraph{Runtime verification of linearizabilty}
As far as we know, runtime verification of linearizability has only been studied in~\cite{ETQ05, ET06},
in asynchronous failure-free models with centralized monitoring systems.
Those papers study I/O refinement, which generalizes linearizability for
objects without sequential specifications. \GenLin includes objects without sequential
specifications too as it includes set-linearizability and  interval-linearizability~\cite{CRR18, CRR23, N94}.
In~\cite{ETQ05, ET06}, specific code is added to a \emph{white-box} implementation 
in order to record the execution in a \emph{log} (a sequence of events)
that later is verified by a single process
(hence the runtime verification algorithm is neither distributed nor fault-tolerant).
Events must be atomically recorded in the log, which necessarily requires 
consensus or the use of locks~\cite{HS08,MS04,R13}. High-level operations
are divided in \emph{mutators} and \emph{observers}.
For mutators, the user has to add code that records in the log when the operation takes 
effect (i.e. its linearization point), and for observers, invocations
and responses are recorded separately. 
It is known that there are linearizable implementations whose linearization points
are \emph{not fixed}~\cite{LV95}, 
hence the approach in~\cite{ETQ05, ET06} is not general.
Finally, it is not explained in~\cite{ETQ05, ET06} what the relation is between the
actual execution of the algorithm under inspection and the execution recorded in the log.

\paragraph{Runtime enforcement}
Runtime enforcement is an extension of runtime verification whose aim is to
evaluate the current execution of a system under inspection, and halt the system
whenever it deviates from a desired property (see survey~\cite{FFM12}).
Runtime enforcement initiated with the \emph{security automata} formalism of Schneider~\cite{S00}.
Our interactive model for distributed runtime verification can be understood as
a distributed version of Schneider's security automata.
As far as we know, so far there have been proposed only a few
distributed runtime enforcement algorithms (e.g.~\cite{BJKZ13, GMS11, GF21, HKBEF18, SGK20}).

\paragraph{Accountability}
In general, accountability requires correct processes to \emph{irrevocably} detect safety violations. 
Note that false positives are not allowed: once a violation is detected, the detection cannot
be revoked. The concept of accountability in the context of distributed computing was introduced in~\cite{HKD07}.
Motivated by blockchain technologies, there have been recently proposed accountable 
algorithms for consensus~\cite{BG17, CGG21, CGGGK22, SWNKV21} and general tasks~\cite{CGGGKMS22}.
All these works consider semi-synchronous message-passing systems with malicious Byzantine failures,
and the main target is to irrevocably detect Byzantine processes.
Here we consider concurrent systems with benign crash failures, hence processes never deviate from its specification.
In this scenario the safety violation one can detect are invalid outputs, 
as our self-linearizable implementations do.

\section{Final Discussion}
\label{sec-final}

This paper studied the problem of distributed runtime verification of linearizability
in asynchronous wait-free shared memory systems, through 
a novel interactive model for runtime verification of correctness conditions.
Distributed runtime verification of linearizability is \emph{not} an agreement problem:
regardless of the consensus number of the base objects used for verification,
 the problem is impossible for common sequential objects 
such as queues, stacks, sets, priority queues, counters and the consensus problem.
However, a predictive version of the problem can be solved
for the class \RV of concurrent implementations, and without the need of consensus. 
Moreover, the possibility result holds
for a correctness condition \GenLin that includes linearizability and generalizations of it such as
set-linearizability~\cite{N94} and interval-linearizability~\cite{CRR18, CRR23},
the latter known to be expressive enough to model tasks~\cite{HKR13}
and any concurrent object satisfying some reasonable assumptions~\cite{CRR18, GLM18}.
\GenLin contains any object that is closed
by prefixes and similarity, the latter being a property identified here.
Any concurrent implementation can be transformed into
its counterpart in \RV, 
and there is a wait-free verifier that satisfies predictive soundness and
completeness, for the class~\RV and any object in \GenLin.
A crucial  building block in the transformation to obtain \RV implementations
is that of the views mechanism for capturing real-time order of executions~\cite{CRR18}.
Read/write objects suffice to solve predictive runtime verification,
hence consensus among two or more processes is not needed.

A simple and generic methodology for designing self-enforced correct
\GenLin implementations was obtained. 
Given any concurrent implementation for some \GenLin object, 
one can produce a self-enforced correct concurrent implementation 
with the same progress properties such that all outputs are 
runtime verified,
or the implementation blocks, reporting error to every new invocation. 
These implementation are able to produce a certificate of the current computation at any time,
hence allowing the design of systems in a modular manner
with accountable and forensic guarantees.
We are not aware of previous concurrent implementations in the literature with such properties.
As far as we know, this is the first distributed runtime verification algorithm, for any correctness condition,
that is at the same time fully asynchronous and fault-tolerant. 

All together, the results show that runtime verification is possible if and only
if the concurrent implementation under inspection outputs some information
of its current history. The methodology for self-enforced correct implementations
show how any \GenLin implementation can be instrumented to make it runtime verifiable.

We believe several directions are worth to be explored. A natural direction is to study other correctness conditions
in our interactive model
such as sequential consistency~\cite{L79} 
or causal consistency~\cite{ANBKH95}.
Runtime verification of hyperproperties~\cite{CS10} is interesting as well.
The main algorithmic technique in our solutions is that of the views mechanism for sketching the current
execution. In general it is interesting to explore if the mechanism helps to runtime verify other properties.
In this direction, we have shown that the views mechanism allows
predictive runtime verification of eventual consistency conditions such as strong eventual counters~\cite{CR25}.

Finally, conducting experimental evaluations are important to understand
if the proposed algorithms can provide good performance in practical settings.
{In follow-up work~\cite{RC24}, we have improved the step complexity of 
$\m{A}^*$ and $\m{V}_\m{O}$ by removing the usage of snapshot objects. 
From the improved algorithms, it is derived a decoupled 
version of the self-enforced \GenLin implementation 
(in the style of $\m{D}_{\m{O}, \m{A}}$ in Section~\ref{sec-decoupled}),
where, in order to produce a response, a producer executes the steps in $\m{A}$ 
plus only \emph{five} additional steps. The aim of those additional steps is capturing and storing the sketch of $\m{A}^*$.
It is yet to be empirically tested the performance of these improved implementations.}

\appendix

\section{Linearizability for Some Objects is not Predictively Verifiable}
\label{sec-impossibility-predict}

\begin{theorem}[Impossibility of Distributed Runtime PredictiveVerification]
\label{theo-impossibility-weak}
Linearizability for queues, stacks, sets, priority queues, counters
and the consensus problem (modeled as 
sequential objects) is not 
distributed runtime predictively verifiable, 
{regardless of the consensus number of base objects used in a verifier.}
\end{theorem}

\begin{proof}
The proof is nearly the same as the proof of Theorem~\ref{theo-impossibility}, 
where executions $E$ and $F$ of a hypothetical wait-free verifier $\m{V}_{\sf queue}$ are obtained 
from the non-linearizable queue implementation~\m{A} defined in the proof.
The difference between the proofs is in the last step, where it is now observed that the execution $F$
can be equally obtained from any wait-free linearizable queue implementation \m{B} 
(several such implementations appear in~\cite{HS08}).
The proof now concludes by observing that processes cannot report \error in $F$ 
(which is allowed by predictive soundness) 
because there is no witness for $\m{B}$ as it is linearizable. Thus, by indistinguishability,
 no process reports \error~in~$E$, and hence $\m{V}_{\sf queue}$  does not satisfy completeness.
Therefore, $\m{V}_{\sf queue}$~cannot~exist.
\end{proof}

\subsection*{Acknowledgements}
We would like to thank Hagit Attiya, Gregory Chockler, Ori Lahav, Sergio Rajsbaum, David Rosenblueth, Serdar Tasiran
and anonymous reviewers for insightful comments and discussions on this work.
Part of this work was done while Armando Casta\~neda was on sabbatical 
leave visiting the  Department of Computer Science of the University of Surrey.
This work was partially supported by the research projects DGAPA-PAPIIT IN108723 and IN103126,
and SECIHTI CBF-2025-I-393.
Valeria Rodr\'iguez-Jim\'enez is the recipient of a PhD fellowship from~SECIHTI.

\bibliographystyle{abbrv}

\end{document}